\documentclass[usenatbib,arxiv]{iupartex}
\usepackage{newtxtext,newtxmath}
\usepackage[T1]{fontenc}
\usepackage{ae,aecompl}
\usepackage{subcaption}

\usepackage{multicol}   
\usepackage{bm}		    

\title[V454 Aurigae]{The Fundamental Parameters and Evolutionary Status of V454 Aurigae}

\author[Y\"{u}cel et al.]{%
G. Y\"{u}cel$^{1\cc}$,\orcid{0000-0002-9846-3788}
R. Canbay$^{2}$\orcid{0000-0003-2575-9892}, and
V. Bak{\i}\c{s}$^{3}$\orcid{0000-0002-3125-9010}
\affsep \\
$^1$Istanbul University, Faculty of Science, Department of Astronomy and Space Sciences, Beyazıt, 34119, Istanbul, Türkiye\\
$^2$Istanbul University, Institute of Graduate Studies in Science, Programme of Astronomy and Space Sciences, 34116, Beyaz{\i}t, Istanbul, Türkiye\\
$^3$Akdeniz University, Faculty of Science, Department of Space Sciences and Technologies, Konyaalt{\i}, 07030, Antalya, Türkiye}

\corres{G. Y\"{u}cel}{gokhannyucel@gmail.com}

\pubyear{2024}

\doiheader{XXXXXXX/PAR.20XX.00000}
\date{
	\pSubmit{00.00.0000} 
	\pRevReq{00.00.0000}
	\pLastRevRec{00.00.0000}
	\pAccept{00.00.0000}
	\pPubOnl{00.00.0000}
}
\volume{0}
\volnumber{1}

\begin{document}
\label{firstpage}
\pagerange{\pageref*{firstpage}--\pageref*{lastpage}}
\maketitle

\setlength{\tabcolsep}{5mm} 
\renewcommand{\arraystretch}{1.0} 

\begin{abstract}
Eclipsing binary systems have a unique feature, which enables scientists to obtain precise fundamental star parameters, which opens to greater area of astrophysics studies. In this study, we have derived the fundamental parameters, the evolutionary status, and the birthplace of V454 Aur in the Galaxy by combining radial velocity, photometric, and spectral energy distribution data. We have updated ephemerides and period of V454 Aur as $2458850.80136_{-0.00001}^{\,+0.00001}$ and $27.0198177_{-0.0000003}^{\,+0.0000003}$, respectively. We obtain $1.173^{\,+0.016}_{-0.016}$ $M_\odot$ and $1.203^{\,+0.022}_{-0.026}$ $R_\odot$ for the primary component and $1.045^{\,+0.015}_{-0.014}$ $M_\odot$ and $0.993^{\,+0.034}_{-0.027}$ $R_\odot$ for the secondary component. The effective temperatures for the components are accurately determined via SED data as $6250^{\,+150}_{-150}$ K and $5966^{\,+109}_{-89}$ K for the primary component and the secondary component, respectively. The metallicity of the components is derived from evolutionary tracks, which implies slightly higher than Solar metallicity. According to analysis, the components of V454 Aur are on the main sequence. Our distance calculation for the system is, which is $65.07^{\,+2}_{-3}$ pc, in excellent agreement with \textit{Gaia} astrometric data, which is $65.07^{\,+0.09}_{-0.09}$ pc. The current age of the system is $1.19_{-0.09}^{\,+0.08}$ Gyr, and it will start to mass transfer between components in 5 Gyr from now on. Dynamical orbital analysis shows that V454 Aur follows a boxy pattern around the Galactic centre and is a member of the thin-disc component of the Galaxy. Considering the age and metallicity of this system, it was found to have formed just outside the Solar circle. 

\end{abstract}

\begin{keywords}
stars:binaries -- stars:fundamental parameters -- stars:evolution -- stars:kinematics
\end{keywords}



\section{Introduction}
In principle, to understand our Galaxy, and therefore the universe, and its evolution, we need to understand the cornerstone of it, which is stars. The evolution of a star is based on its mass, which is the dominant one, and its chemical composition. There are several ways to acquire the mass of a star, but among them, eclipsing binaries are the most accurate ones \citep{Serenelli}. Eclipsing binaries, in general, are the centre of astrophysics studies since they provide valuable information (mass, radius, temperature, etc.) of the component stars within 1-3\% uncertainty rates, or mostly even better according to the quality of the data that are used in analysis \citep{Torres, Prsa2020}. Eclipsing binaries with known fundamental parameters are increasing every day, but this is not a reason to stop investigating new eclipsing binary systems and obtaining their parameters, since every known system is a source for several studies from stellar populations \citep[e.g.][]{Chabrier, Moe} to empirical MLR studies \citep[e.g.][]{Benedict, Eker2015, Eker2018, Eker2024}. Therefore, there is still a need to study eclipsing binaries and derive their fundamental parameters precisely.

V454 Aur (HD 44192, SAO 59016, Hip 30270, $l=178^{\rm o}.803546,~~b=09^{\rm o}.510553$), is a Northern Hemisphere, detached eclipsing binary system. The variable star feature of V454 Aur was discovered by \textit{Hipparcos} \citep{Perryman}. The first ground-based observation of V454 Aur was done by \cite{Griffin2001} via obtaining photoelectric radial velocities (RVs). As a result, the spectroscopic orbit of V454 Aur was calculated. Later, \cite{Nordstrom} calculated the temperature, metallicity, and age of the star as 6025 K, -0.14 dex, and 5 Gyr, respectively, by using \textit{ubvy} photometric data. These values were improved by \cite{Holmberg} and later on by \cite{Casagrande} for temperature, $\log{g}$, metallicity, and age as 6064 K, -0.08 dex, 4.43 dex, and 4 Gyr, respectively.  
\cite{PrsaTESS} have used \textit{TESS} \citep{Ricker} observations to calculate ephemerides and period of V454 Aur and obtained the values of 2458850.778358$\pm$0.002306 and 17.8883306$\pm$0.0064858 days, respectively. However, they did the analysis based on only 20 sector. There is no prior study of this system.

In this study, we have combined radial velocity and photometric data with multiple sectors, which is provided by \textit{TESS}, and obtained fundamental parameters of V454 Aur for the first time in the literature. We have studied its evolution scenarios and found the system's initial orbital parameters and kinematics of the system, which tell us where this system was born. 

This paper is structured as follows. In Section \S2, we present the properties of observational data used. In Section \S3, the calculated fundamental parameters of the system are presented. In Section \S4, we present the detailed evolutionary analysis. In Section \S5, kinematics of the system. Finally, in Section \S6, we have discussed our comprehensive results.

\section{Data}

\subsection{Radial Velocities}

Radial velocity data (RVs), which were used in this study, have been taken from \cite{Griffin2001}. Details can be found in that paper, but we briefly give a summary here. \cite{Griffin2001} observed V454 Aur in the years of 2000-2001, with a total of 65 observations for both components at Observatoire de Haute-Provence\footnote{\url{http://www.obs-hp.fr}} with
1-m Swiss telescope, equipped with Coravel \citep{Baranne} instrument. The RVs that were used in this study are given in Table \ref{tab:RVs}.

\begin{table*}[ht]
  \centering
    \caption{Journal of spectroscopic observations.}
  \resizebox{\textwidth}{!}{
  \begin{tabular}{lcccccccc}
  \hline
    HJD-2400000 & RV$_1$ & RV$_2$ & HJD-2400000 & RV$_1$ & RV$_2$ & HJD-2400000 & RV$_1$ & RV$_2$ \\ 
     & (km s$^{-1}$) & (km s$^{-1}$) &  & (km s$^{-1}$) & (km s$^{-1}$) &  & (km s$^{-1}$) & (km s$^{-1}$) \\ 
\hline
    51595.994 & -29.3 & -52.5 & 51852.055 & -54.1 & -24.5 & 51906.102 & -54.1 & -24.6 \\
    51602.935 & -76.8 & 0.0 & 51852.097 & -53.6 & -27.0 & 51906.115 & -54.6 & -25.0 \\
    51604.912 & -93.2 & 20.0 & 51852.149 & -52.6 & -26.3 & 51906.128 & -54.3 & -24.9 \\
    51606.817 & -95.2 & 22.9 & 51852.199 & -51.5 & -29.5 & 51906.141 & -54.8 & -24.0 \\
    51607.009 & -92.9 & --  & 51852.253 & -50.8 & -31.2 & 51908.029 & -21.2 & -64.3 \\
    51607.936 & -76.0 & 0.6 & 51861.198 & -11.1 & -73.4 & 51908.894 & -13.2 & -72.4 \\
    51609.912 & -33.1 & -48.2 & 51863.154 & -16.7 & -66.5 & 51916.051 & -13.0 & -71.2 \\
    51624.908 & -38.9 & -42.0 & 51865.156 & -25.4 & -58.7 & 51917.973 & -19.3 & -64.2 \\
    51627.947 & -60.2 & -18.1 & 51870.143 & --   & -27.3 & 51918.947 & -24.4 & -59.6 \\
    51628.889 & -67.8 & -9.8 & 51870.225 & -- & -25.5 & 51920.099 & -29.4 & -54.0 \\
    51639.886 & -7.0 & -77.0 & 51870.268 & -54.0 & -25.7 & 51922.976 & -44.6 & -36.0 \\
    51640.896 & -5.2 & -79.8 & 51878.126 & -77.0 & 0.6 & 51924.956 & -58.6 & -19.9 \\
    51641.857 & -5.4 & -79.4 & 51880.176 & -31.7 & -47.5 & 51925.961 & -67.1 & -10.8 \\
    51657.870 & -85.8 & 10.3 & 51881.096 & -19.1 & -62.7 & 51926.926 & -74.4 & -4.0 \\
    51660.892 & -95.1 & 21.0 & 51887.089 & -7.6 & -76.2 & 51934.972 & -21.8 & -62.7 \\
    51812.171 & -29.4 & -51.4 & 51892.066 & -23.7 & -58.7 & 51936.957 & -7.8 & -77.3 \\
    51823.191 & -93.3 & 18.8 & 51900.062 & -76.2 & -0.8 & 51946.977 & -27.9 & -53.8 \\
    51826.158 & -32.3 & -50.8 & 51906.057 & -55.6 & -23.8 & 51954.959 & -84.2 & 8.5 \\
    51834.181 & -10.9 & -72.6 & 51906.068 & -55.4 & -24.0 & 51956.851 & -97.7 & 24.7 \\
    51851.043 & -78.2 & 1.7 & 51906.080 & -55.0 & -23.8   & 51981.946 & -83.7 & 8.5 \\
    51851.972 & -57.0 & -22.0 & 51906.091 & -55.3 & -24.6 &       &       &  \\
    \hline
    \end{tabular}}%
  \label{tab:RVs}%
\end{table*}%

\subsection{Photometric Data}

Photometric data, which are used in this study, have been obtained via \textit{TESS}. Although the main purpose of \textit{TESS} is that of finding exoplanets by looking at brightness changes of a star, it also has been a source of producing light curves of eclipsing binaries, which are also needed to analyze eclipsing binary systems \citep{PrsaTESS}.

\textit{TESS} has observed V454 Aur in a total of five sectors, which are 20, 43, 44, 45, and 60 with exposure times 1800s, 600s, 600s, 600s, and 200s, respectively. \textit{TESS} has also observed V454 Aur on sectors 71, 72, and 73, but those photometric data are not available yet. 

\begin{figure*}[ht]
\centering
\includegraphics[width=.49\textwidth]{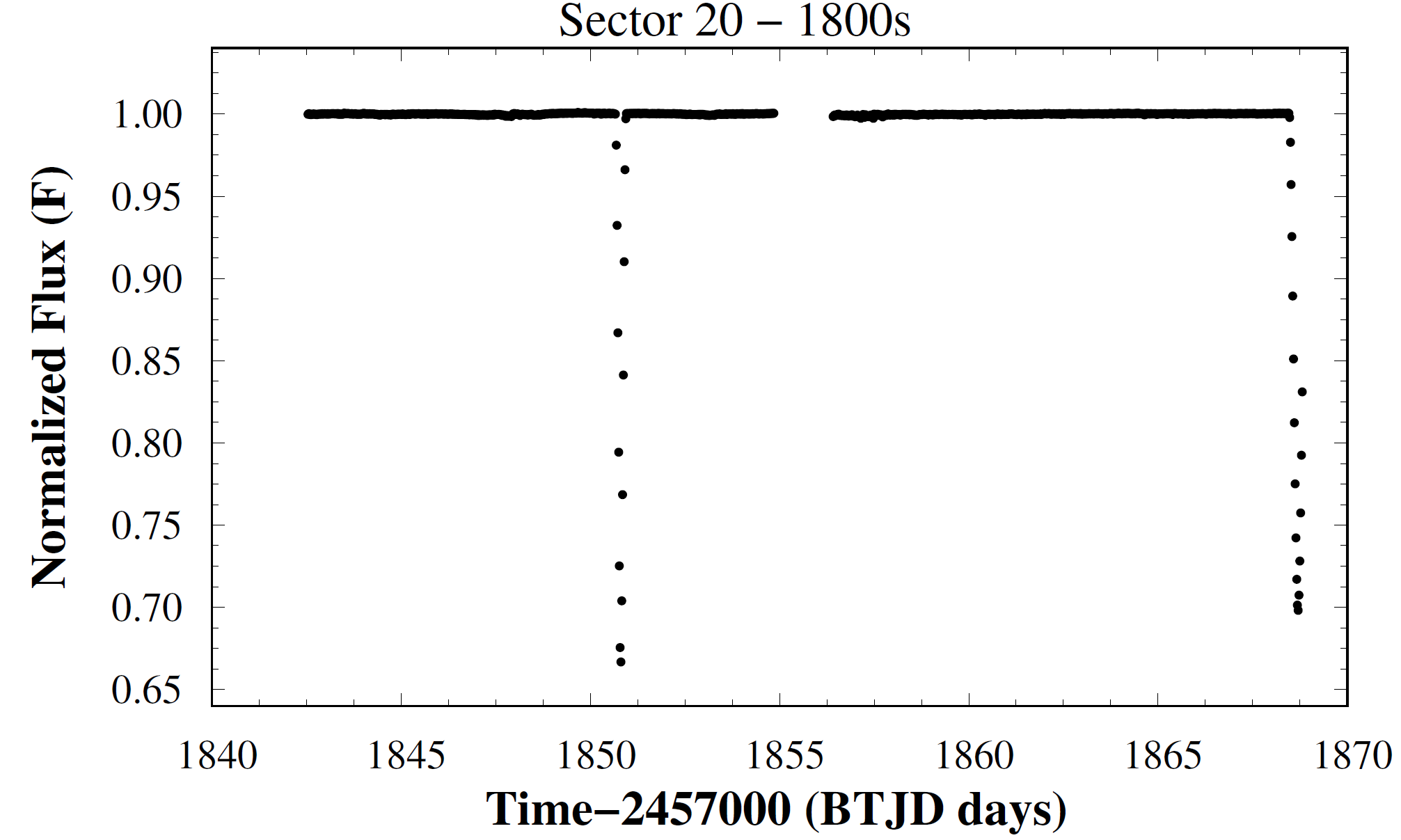}
\includegraphics[width=.49\textwidth]{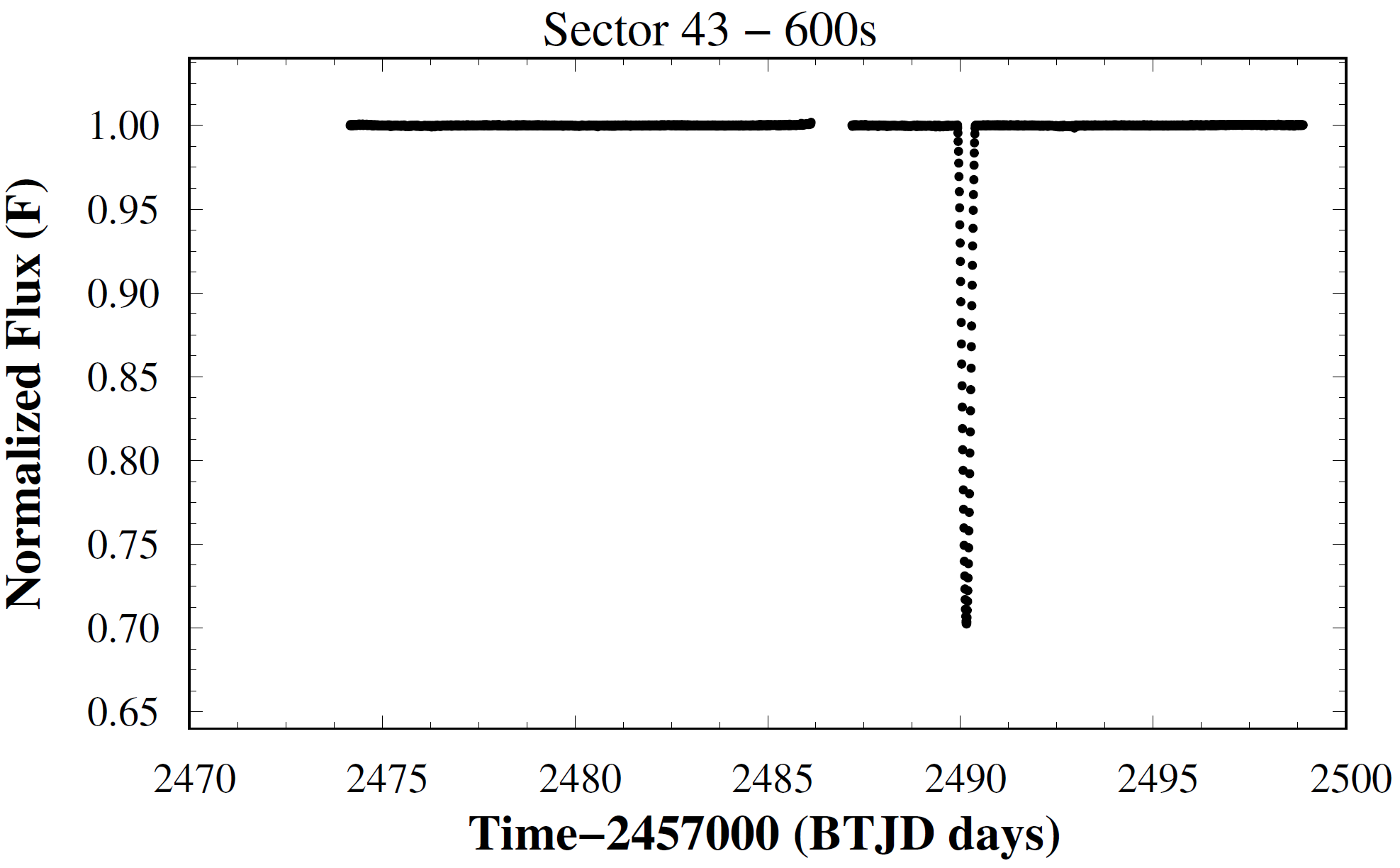}\medskip

\includegraphics[width=.49\textwidth]{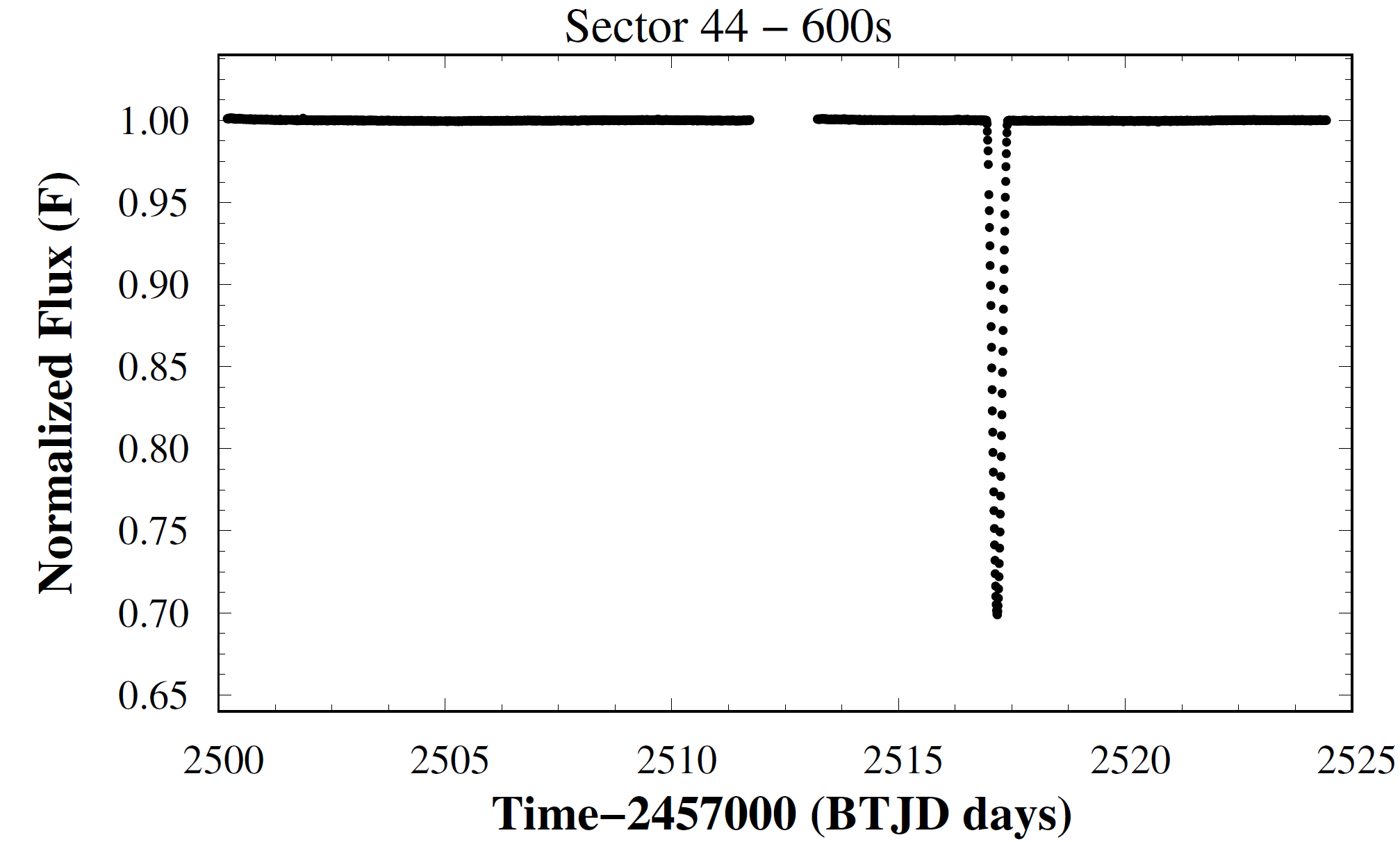}
\includegraphics[width=.49\textwidth]{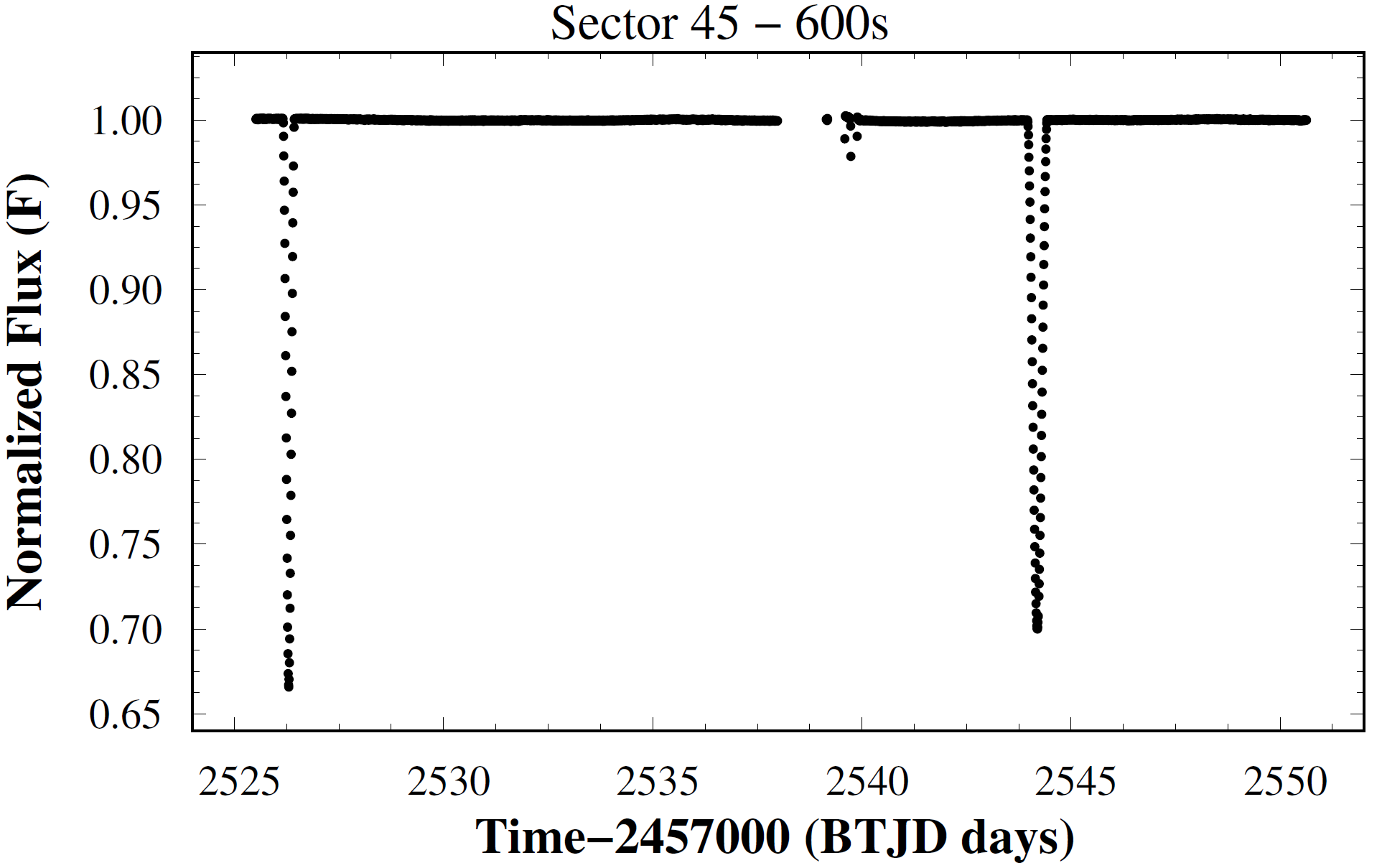}

\medskip
\includegraphics[width=.5\textwidth]{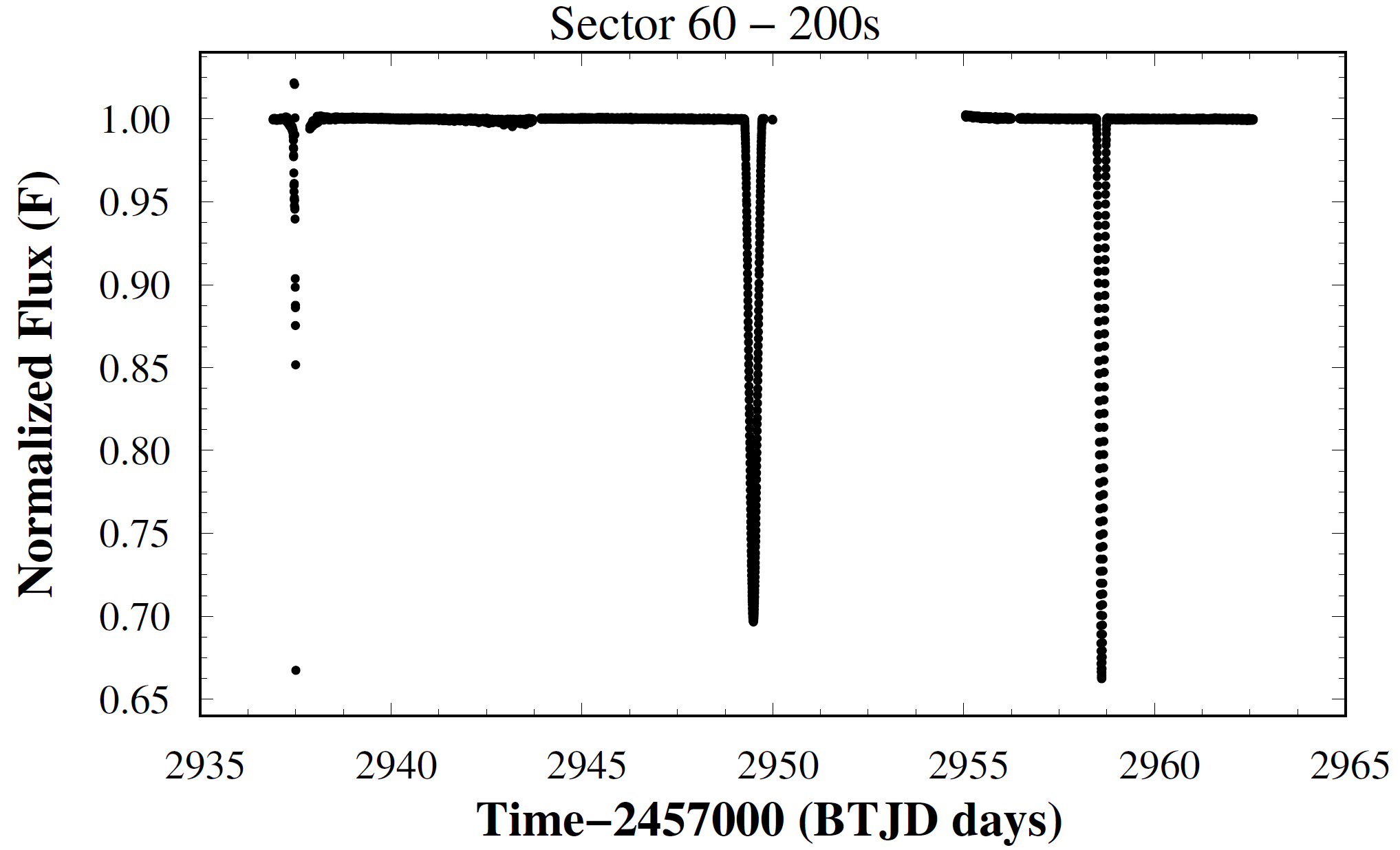}

\caption{\textit{TESS} observation of V454 Aur in five different sectors.}
\label{Tess-images}
\end{figure*}

We have used \texttt{Lightkurve} v2.4 \citep{lk} to acquire photometric data. 200s photometric data were used in analysis wherever available, but missing parts were completed by other sectors. The photometric data used in this study are shown in Figure~\ref{Tess-images} by each sector.

\newpage

\section{Fundamental Parameters}

\subsection{Analysis of Ephemerides}

\begin{figure*}[ht]
\centering
\includegraphics[width=\textwidth]{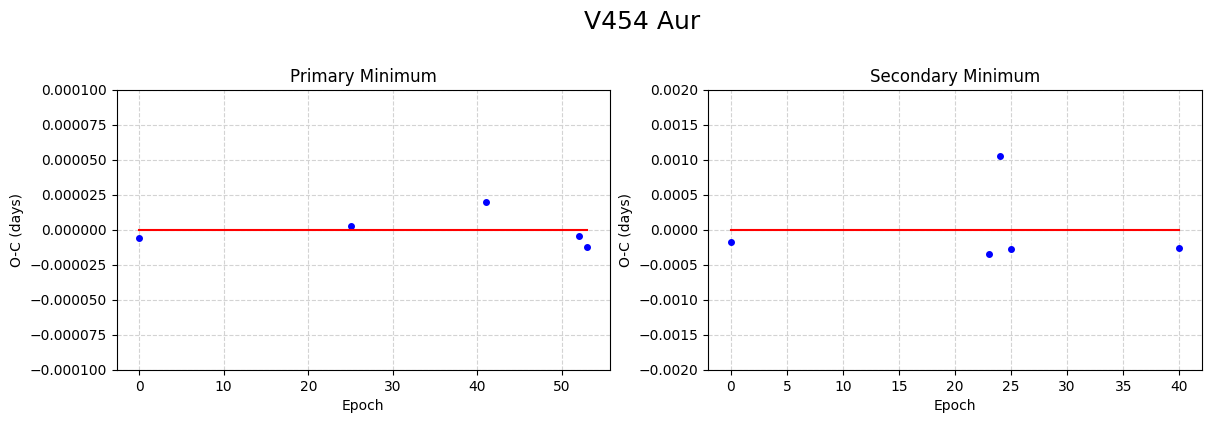}
\caption{O--C diagram for primary and secondary minima.}
\label{fig.o_c}
\end{figure*}

In the literature, there is only one time of minimum measurement \citep[{\it Hipparcos},][]{Perryman} for the system, which makes its long-term orbital period change study impossible. Nevertheless, {\it TESS} satellite obtained five consecutive measurements (sectors; see Figure \ref{Tess-images}) and using {\it TESS} measurements allow us to determine the up-to-date ephemeris of the system. We have measured the times of minima available in the \textit{TESS} photometric data using the Kwee-Van Woerden method \citep{Kwee}, which are presented in Table \ref{tab:ToM}. Since the system has an eccentric orbit, primary and secondary ephemerides are calculated separately. The linear least-squares method to the $O-C$ data ($O-C=T-(T_0+P \times E$)) yielded the following ephemerides:

\begin{equation}
	T ({\rm HJD}) = 2458850.80136 (1) +  27.0198177 (3) \times E
\end{equation}

for the primary minimum,

\begin{equation}
	T ({\rm HJD}) = 2458868.69867 (61) + 27.0198105 (233) \times E
\end{equation}

for the secondary minimum. The O--C diagram is shown in Figure \ref{fig.o_c}.

\begin{table}[ht]
	\centering
	\caption{The times of minima of V454 Aur extracted from \textit{TESS} data.}
		\begin{tabular}{cc}
			\hline
			Times of minima & Component \\
			(HJD)           &           \\
			\hline
			2458850.80135	&	pri	\\
			2458868.69850	&	sec	\\
			2459490.15397	&	sec	\\
			2459517.17517	&	sec	\\
			2459526.29680	&	pri	\\
			2459544.19366	&	sec	\\
			2459949.49083	&	sec	\\
			2459958.61390	&	pri	\\
			2460255.83187	&	pri	\\
			2460282.85168	&	pri	\\
			\hline
	\end{tabular}
	\label{tab:ToM}%
\end{table}%

\subsection{Spectroscopic Orbit and Light Curve Modelling}

Although the spectroscopic orbit of V454 Aur has been calculated by \cite{Griffin2001}, the system has been reanalyzed with simultaneous solutions using both RVs data and photometric data to obtain fundamental parameters such as mass, radius, temperature ratio, and the light contributions of both components via using PHysics Of Eclipsing BinariEs v1.0\footnote{\url{http://phoebe-project.org/1.0/download}} \citep[PHOEBE,][]{phoebe} which is based on Wilson-Devinney code \citep[WD,][]{wd1,wd2,wd3,wd4,wd5}. 

Since there is no indication in photometric data of any mass transfer in the system, the analysis has been made in detached mode. During the analysis, conjunction time $T_0$, orbital period $P$, and the temperature of the primary component have been fixed (see the next section for how the primary component's temperature has been determined) and the following parameters have been adjusted: mass ratio ($q$), eccentricity ($e$), the argument of periapsis ($w$), semi-major axis ($a$), systemic velocity ($V_\gamma$), the orbital inclination ($i$), the temperature of secondary component ($T_2$), dimensionless surface potentials of both components ($\Omega_{1,2}$), and monochromatic luminosity ($L_1$). As the limb-darkening (LD) calculations, we have adopted logarithmic LD laws. LD values calculated from a Fortran code written by Walter van Hamme\footnote{\url{https://faculty.fiu.edu/~vanhamme/ldfiles/ldinterpol.for}}.

\begin{figure}
\centering
	\includegraphics[width=0.9\linewidth]{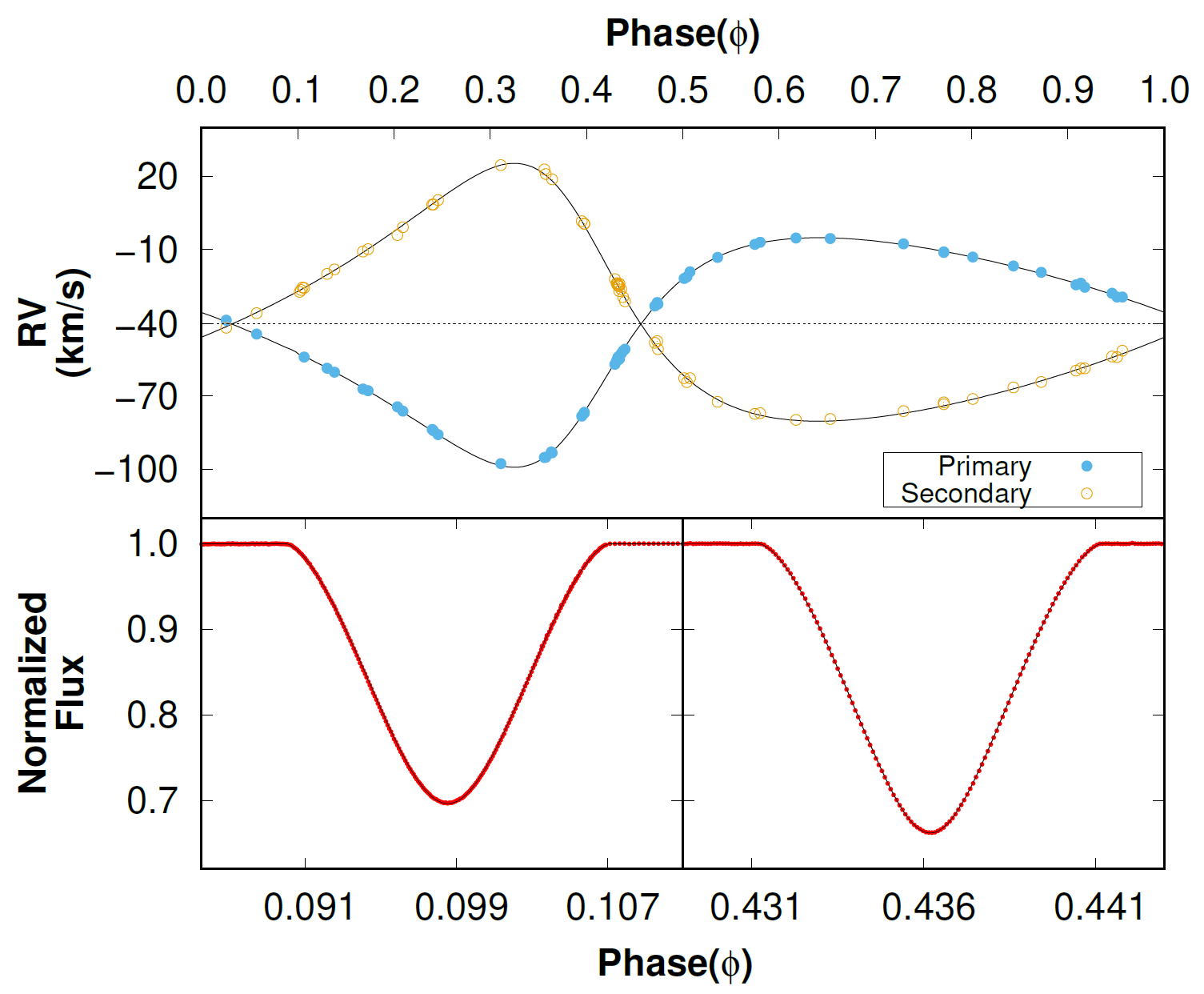}
     \caption{Observed RVs with the best fitting radial velocity curves and photometric data with LC modeling. Blue and yellow circles represent the RVs of the primary and secondary components of V454 Aur, respectively. In LC modeling, red dots represent the photometric data and the black curve is the best LC model.}
    \label{fig:rvcurve}
\end{figure}

After the initial analysis, we used a custom Markov chain Monte Carlo (MCMC) sampler\footnote{\url{https://sourceforge.net/p/phoebe/mailman/message/33650955/}} based on \texttt{emcee} \citep{mcmc} to improve the parameters of components of V454 Aur and to determine the heuristic errors. The sampler ran with 128 walkers and 1000 iterations as conventional. LC and spectroscopic orbit models are presented in Figure~\ref{fig:rvcurve} and the fundamental parameters and heuristic errors for V454 Aur are presented in Table~\ref{table:fund}, as well.

\subsection{Component Temperatures}
\begin{figure}
\centering
	\includegraphics[width=0.9\linewidth]{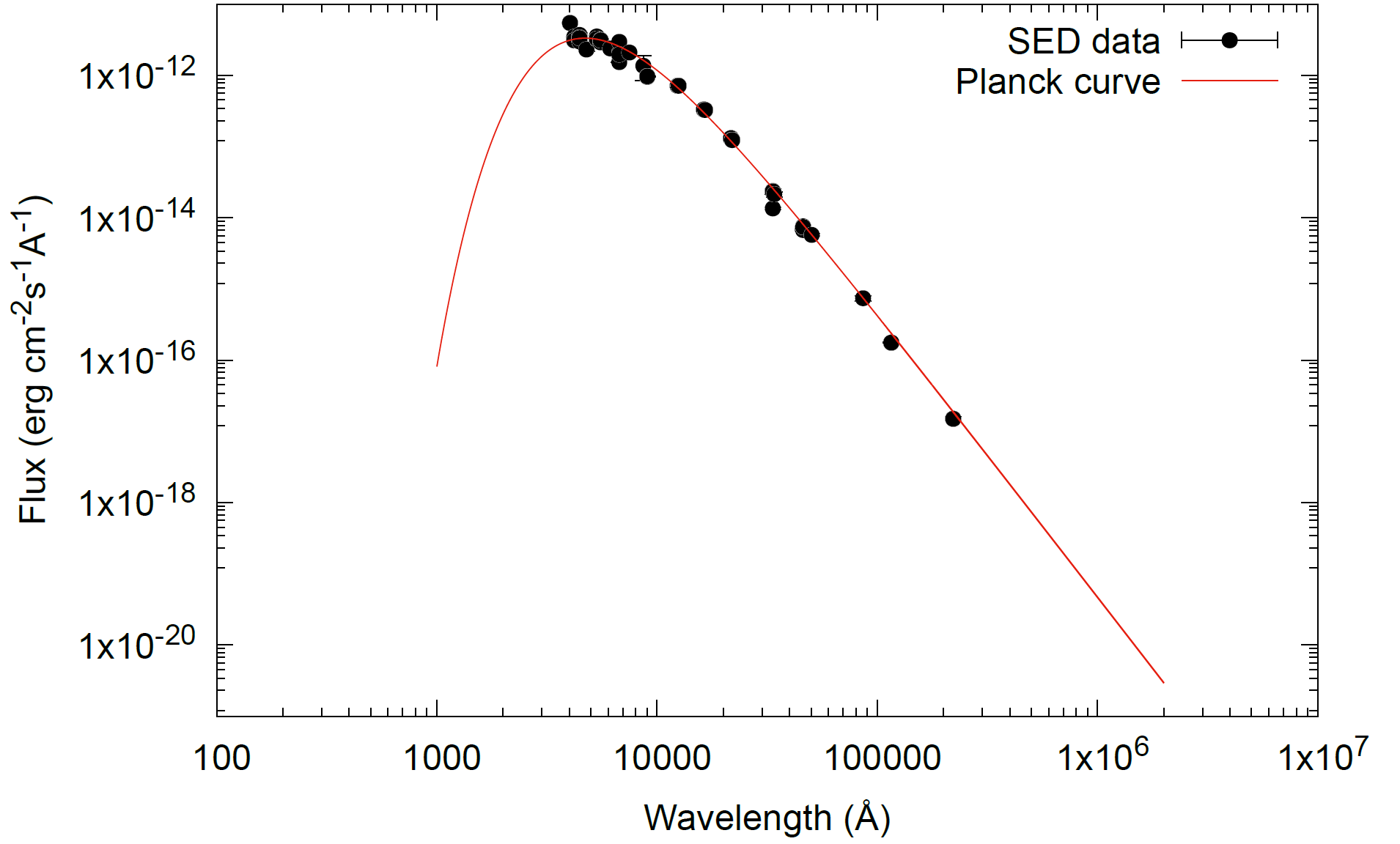}
     \caption{SED data and Planck curve with binary parameters. The figure axes are in log scale. The SED data corresponds to within 1 arcsec around V454 Aur.}
    \label{fig:sedcurve}
\end{figure}

\begin{figure*}
    \centering
	\includegraphics[width=\textwidth]{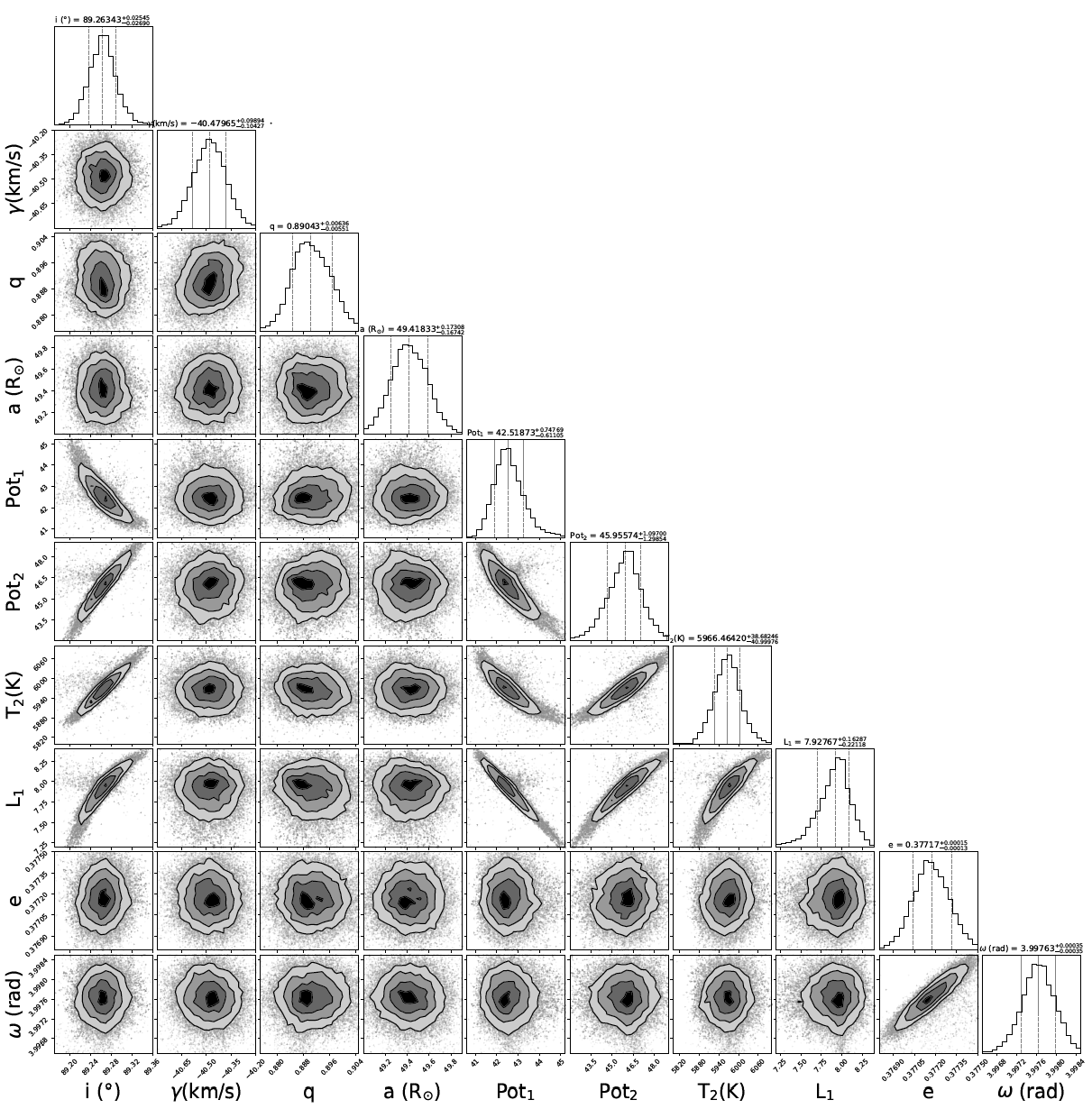}
     \caption{A corner plot of the posteriors for the fundamental parameters of V454 Aur.}
    \label{fig:posterior}
\end{figure*}

The \textit{Gaia} DR3 trigonometric parallax of V454 Aur is $\varpi=15.3669\pm0.0217$ mas corresponding to a distance of $65.07\pm0.09$ pc. At this distance, it is expected to be a very small interstellar extinction. Therefore, the observed $B-V$ colour indicates an extinction-free colour and helps to have a preliminary temperature estimation. The observed $B-V$ of V454 Aur is given to be 0.57 mag corresponding to a temperature of 5879 K using the colour-$T_{\rm eff}$ calibration table in \cite{Bakis2022}. It should be noted that this colour is the combined colour of two components, making the primary component seem cooler. To make a more reliable temperature estimation of the components, we obtained the SED data of V454 Aur and modelled it with the Planck curve as described in  \cite{Bakis2022}. While modelling the SED data, the temperature ratio obtained from the LC analysis, the absolute radii of the components, and the distance to V454 Aur are fixed. In Figure~\ref{fig:sedcurve}, we show the SED data of V454 Aur and the best-fitting Planck curve. The temperature of the components is determined to be 6250 K and 5966 K for primary and secondary components, respectively. The corner plot of the posterior distribution of the fundamental parameters of V454 Aur is  presented in Figure~\ref{fig:posterior}.

\renewcommand{\arraystretch}{1.5}
\begin{table}
\small
\caption{Binary stellar parameters and heuristic errors of V454~Aur.} \label{tab:absolutepar}
\begin{tabular}{lccc}\hline
Parameter & Symbol  & Primary & Secondary \\
\hline
Spectral type & Sp & F1 V-IV & F1 V-IV \\
Ephemerides time (d) & \emph{T}$_{\rm 0}$ & \multicolumn{2}{c}{$2458850.80136_{-0.00001}^{\,+0.00001}$} \\
Orbital period (d) & \emph{P} & \multicolumn{2}{c}{$27.0198177_{-0.0000003}^{\,+0.0000003}$} \\
Mass ($M_\odot$) & \emph{M} & $1.173_{-0.016}^{\,+0.016}$ & $1.045^{\,+0.015}_{-0.014}$ \\
Radius ($R_\odot$) & \emph{R} & $1.203_{-0.026}^{\,+0.022}$ & $0.993_{-0.027}^{\,+0.034}$\\
Surface gravity (cgs) & $\log g$ & $4.345_{-0.022}^{\,+0.025}$ & $4.465_{-0.035}^{\,+0.031}$ \\
Separation ($R_\odot$) & \emph{a} & \multicolumn{2}{c}{$49.418_{-0.167}^{\,+0.173}$} \\
Orbital inclination ($^{\circ}$) & \emph{i} & \multicolumn{2}{c}{$89.263_{-0.027}^{\,+0.025}$} \\
Mass ratio & \emph{q} & \multicolumn{2}{c}{$0.890_{-0.005}^{\,+0.006}$} \\
Eccentricity & \emph{e} & \multicolumn{2}{c}{$0.37717_{-0.00013}^{\,+0.00016}$} \\
Argument of perigee (rad) & \emph{w} & \multicolumn{2}{c}{$3.99763_{-0.00035}^{\,+0.00035}$} \\
Light Ratio (\textit{TESS}) & $l/l_{\rm{total}}$ & $0.631_{-0.018}^{\,+0.013}$ & $0.369_{-0.013}^{\,+0.018}$ \\
Temperature (K) & $T_{\rm eff}$ & $6250_{-0.150}^{\,+0.150}$ & $5966_{-0.089}^{\,+0.109}$ \\
Luminosity ($L_\odot$) & $\log$ \emph{L} & $0.297_{-0.061}^{\,+0.057}$ & $0.050_{-0.056}^{\,+0.055}$ \\
Metallicity   & \emph{z} & \multicolumn{2}{c}{$0.017_{-0.002}^{\,+0.002}$} \\
Combined visual magnitude$^1$ & \emph{TESS} & \multicolumn{2}{c}{$7.131_{-0.006}^{\,+0.006}$} \\
Individual visual magnitude  & \emph{TESS$_{1,2}$} & $7.631_{-0.022}^{\,+0.022}$ &
$8.213_{-0.038}^{\,+0.038}$  \\
Combined visual magnitude & \emph{V} & \multicolumn{2}{c}{$7.65_{-0.01}^{\,+0.01}$} \\
Individual visual magnitude & \emph{V$_{1,2}$} & $8.129_{-0.025}^{\,+0.025}$ &
$8.768_{-0.040}^{\,+0.040}$  \\
Combined colour index (mag) & $B-V$ & \multicolumn{2}{c}{$0.57_{-0.03}^{\,+0.03}$} \\
Colour excess (mag) & $E(B-V)$ & \multicolumn{2}{c}{0} \\
Bolometric magnitude  & $M_{\rm bol}$ & $4.010_{-0.167}^{\,+0.163}$ & $4.625_{-0.223}^{\,+0.220}$ \\
Absolute visual magnitude & $M_{\rm V}$  & $4.025_{-0.157}^{\,+0.153}$ & $4.676_{-0.215}^{\,+0.211}$ \\
Bolometric correction (mag)$^2$ & \emph{BC$_{\rm TESS}$} & $0.440_{-0.005}^{\,+0.006}$ & $0.490_{-0.007}^{\,+0.009}$ \\
Bolometric correction (mag)$^3$ & \emph{BC$_{\rm V}$} & $0.093_{-0.008}^{\,+0.009}$ & $0.083_{-0.006}^{\,+0.011}$ \\
Systemic velocity (km\,s$^{-1}$) & $v_{\gamma}$ & \multicolumn{2}{c}{$-40.480_{-0.104}^{\,+0.099}$} \\
Computed synchronization velocity (km\,s$^{-1}$)& $v_{\rm synch}$ & $2.1_{-0.1}^{\,+0.1}$ & $1.9_{-0.1}^{\,+0.1}$ \\
Age (Gyr) & \emph{t} & \multicolumn{2}{c}{$1.19_{-0.09}^{\,+0.08}$} \\
Distance (pc) & \emph{d} & \multicolumn{2}{c}{$65_{-3}^{\,+2}$} \\
{\it Gaia} distance (pc) & \emph{d} & \multicolumn{2}{c}{$65.07\pm 0.09$}\\
\hline
\footnotesize{$^1$\cite{Paegert2022},$^2$\cite{Eker2023},$^3$\cite{Bakis2022}}
\label{table:fund}
\end{tabular}
\end{table}
\renewcommand{\arraystretch}{1}

\newpage
\section{Evolutionary Analysis}

One of the scopes of this study is finding the evolution scenario for V454 Aur using the fundamental parameters we obtained. In this aspect, we have used the version of r23.05.1 of Modules for Experiments in Stellar Astrophysics
\citep[MESA][]{Paxton2011, Paxton2013, Paxton2015, Paxton2018, Paxton2019, Jermyn2023} and MESA SDK v23.7.3 \citep{SDK} to calculate the evolution of V454 Aur. \texttt{MESA} uses a bunch of microphysics data created by several researchers. The MESA EOS is a blend of the OPAL \citep{Rogers2002}, SCVH \citep{Saumon1995}, FreeEOS \citep{Irwin2004}, HELM \citep{Timmes2000}, PC \citep{Potekhin2010}, and Skye \citep{Jermyn2021} EOSes. Radiative opacities are primarily from OPAL \citep{Iglesias1993, Iglesias1996}, with low-temperature data from \citet{Ferguson2005} and the high-temperature, Compton-scattering dominated regime by \citet{Poutanen2017}.  Electron conduction opacities are from \citet{Cassisi2007} and \citet{Blouin2020}. Nuclear reaction rates are from JINA REACLIB \citep{Cyburt2010}, NACRE \citep{Angulo1999} and additional tabulated weak reaction rates \citet{Fuller1985, Oda1994, Langanke2000}. Screening is included via the prescription of \citet{Chugunov2007}. Thermal neutrino loss rates are from \citet{Itoh1996}, Roche lobe radii in binary systems are computed using the fit of \citet{Eggleton1983}.  Mass transfer rates in Roche lobe overflowing binary systems are determined following the prescription of \citet{Ritter1988}, and so on. 

The evolution scenario of V454 Aur was studied in two sections: single-star evolution and binary-star evolution.

\subsection{Single-star evolution}

\begin{figure}[htp]

\centering
\includegraphics[width=0.7\columnwidth]{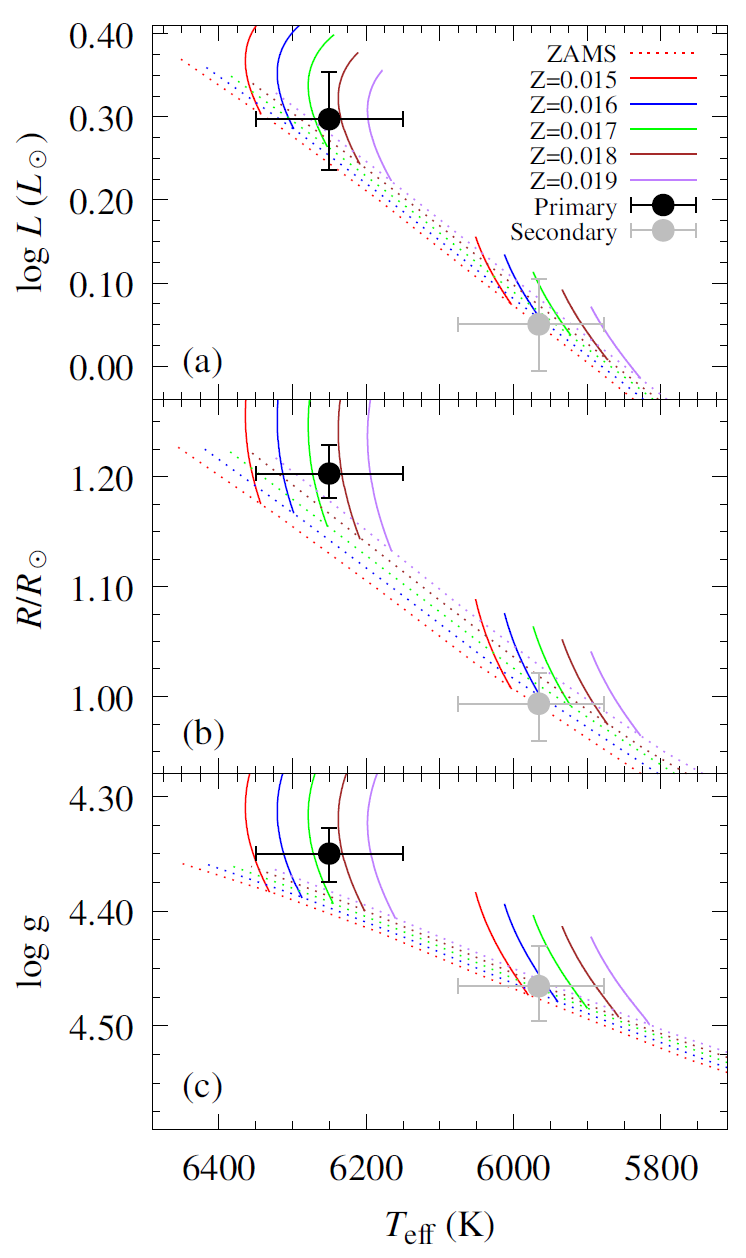}

\caption{Single star evolutionary tracks with different metallicities for V454 Aur on (a) $T_{\rm eff}$--$\log{L}$ plane, (b) $T_{\rm eff}$--$R$ plane, and (c)  $T_{\rm eff}$--$\log g$ plane, respectively.}
\label{fig:single}

\end{figure}

To establish a well-calculated evolution scenario for a star or a system, the metallicity of the star/system has to be known and calculations should be done according to it. Since there is no spectroscopic data for V454 Aur, we have chosen evolution tracks to determine the metallicity of the systems. Eclipsing binary systems were born in the same stellar nurseries/associations, hence, the metallicity of the components of a binary system has to be the same/identical. Considering the system is detached and there is no mass transfer between components, the components can be treated as individual stars. By using \texttt{MESA}, by including the calculated mass of the components of V454 Aur, we built a grid of evolutionary tracks with different metallicities, and ZAMS lines, which are dashed lines. Our results are given in Figure~\ref{fig:single} as $T_{\rm eff}-\log{L}$, $T_{\rm eff}-R$, and $T_{\rm eff}-\log{g}$ planes. Hence, the metallicity ($z$) of the system was determined as $z=0.017\pm0.002$. Both components of V454 Aur are on the main sequence and still burning hydrogen in their cores. 

\subsection{Binary-star evolution}

The components of V454 Aur are on the main-sequence and the system is detached. Hence, the calculation of the evolution of the system with single-star evolution is agreeable. Nonetheless, to understand the evolution of the system from its formation to its end, it's necessary to evolve the system in binary form. \texttt{MESA} enables this option with its \texttt{binary} module. As a starting point, the initial orbital conditions have to be calculated. In this regard, we have used a similar approach that already has been used in literature \citep{Rosales, Soydugan, Gokhan}. We have built a grid with starting different initial orbital periods and different initial orbital eccentricities and ran the evolution with each model until orbital eccentricity dropped to the current eccentricity value of V454 Aur. Then, a $\chi^2$ calculation was made using the determined orbital period of the system, calculated radii, and temperature of the components with every model in the grid. In our calculations, we have activated the option of magnetic breaking \citep{magnetic}, since the components of V454 Aur have convective atmospheres. For the tidal synchronization, we used the ``Orb\_period'' option, which synchronizes the orbit relevant to the timescale of the orbital period. We also applied tidal circularisation, given by \cite{Hurley}.

\begin{figure}
	\includegraphics[width=\columnwidth]{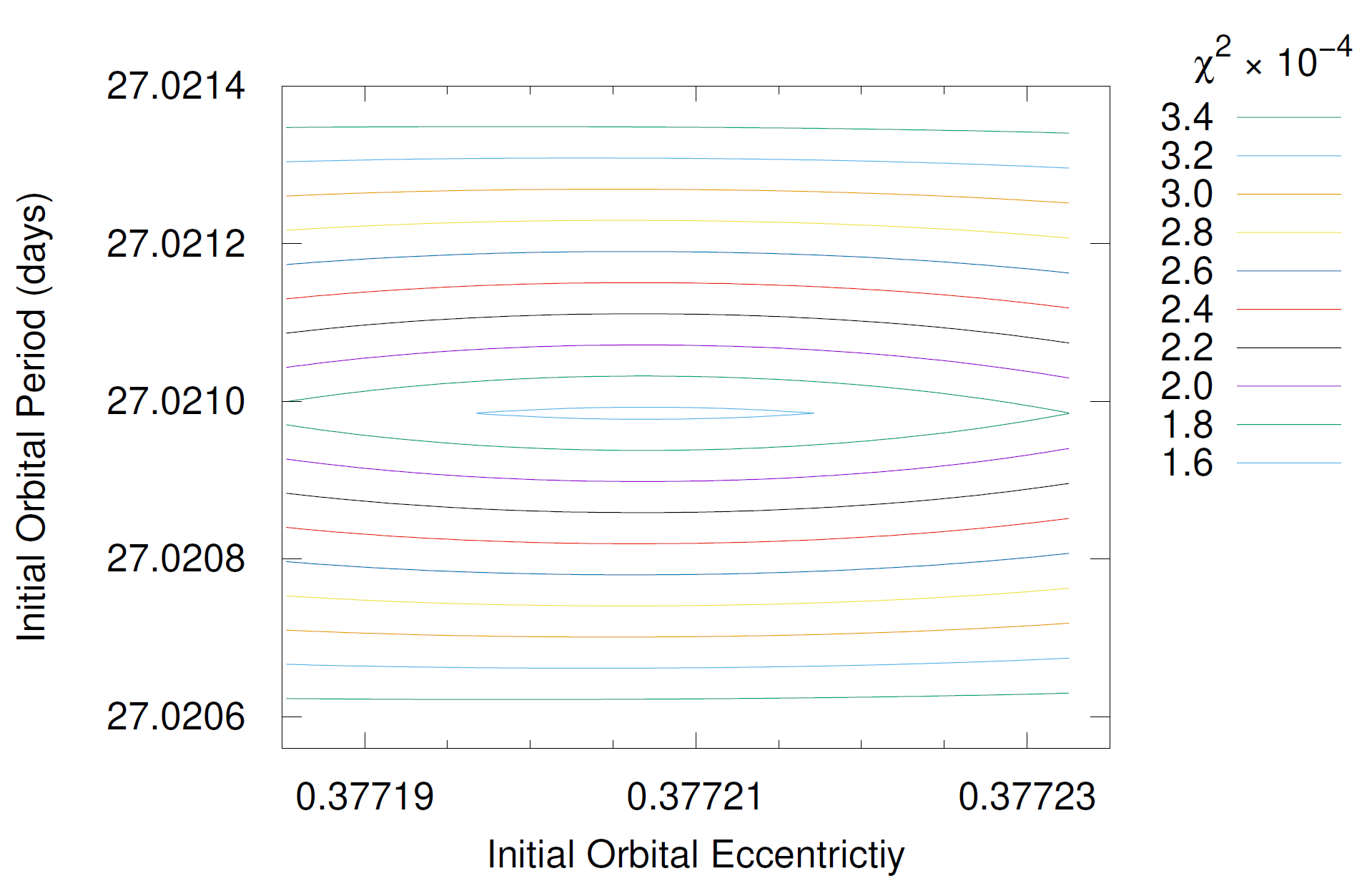}
     \caption{$\chi^2$ map of the corresponding initial period for each initial orbital eccentricity of V454 Aur.}
    \label{fig:chisq}
\end{figure}

In the grid, models that the initial parameters change for the period and eccentricity between 27.020 and 27.040 d with an interval of 0.001 d and between 0.377190 and 0.377300 with an interval of 0.000005, respectively, and kept the evolution continued until eccentricity drops to 0.377170, which is the up-to-date eccentricity value of V454 Aur. The best model, gives the lowest $\chi^2$, 0.00016, with initial orbital parameters for period and eccentricity as 27.021 and 0.377210, respectively (given in Figure~\ref{fig:chisq}).

\begin{figure}
	\includegraphics[width=\columnwidth]{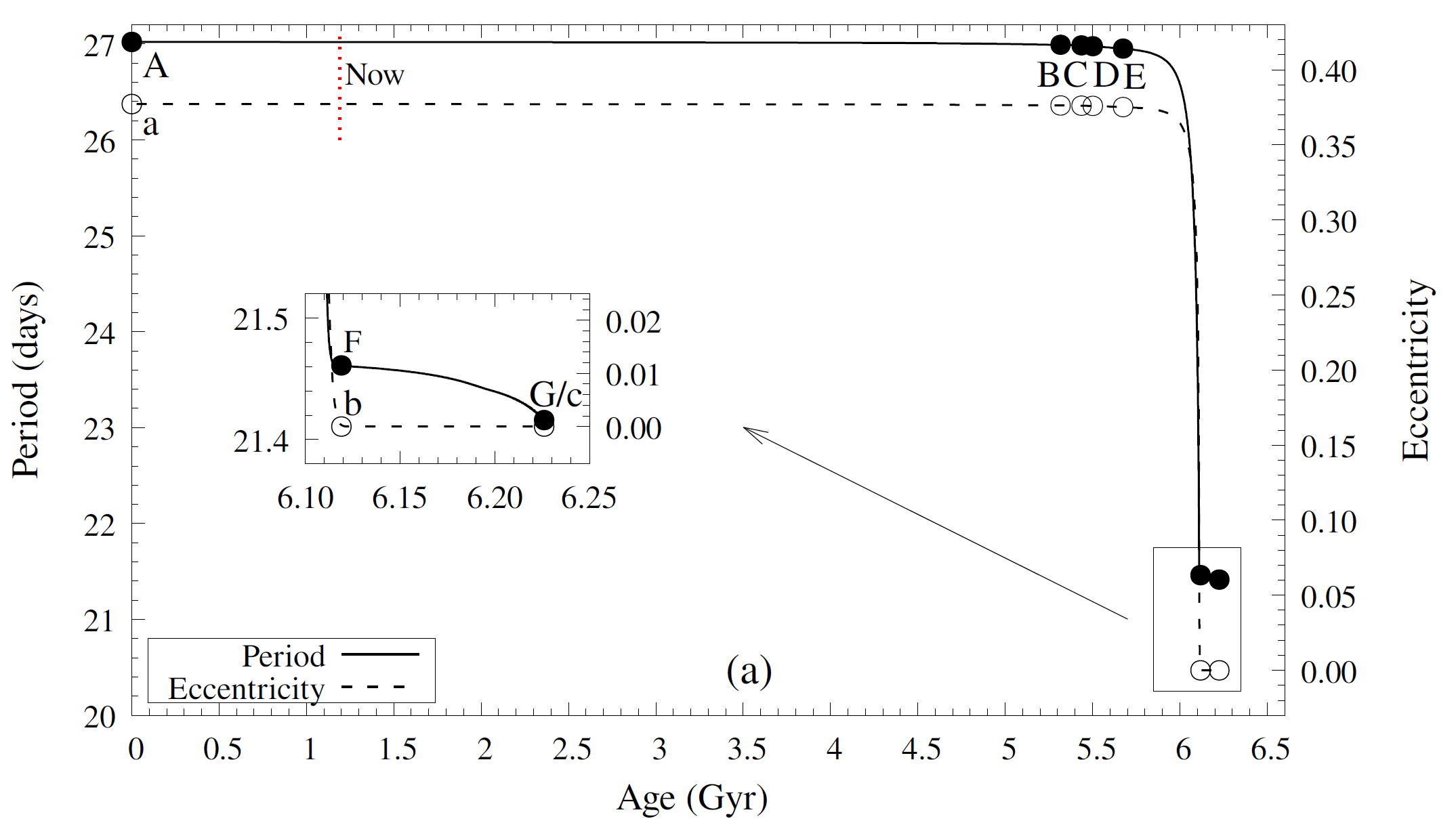}
 \includegraphics[width=\columnwidth]{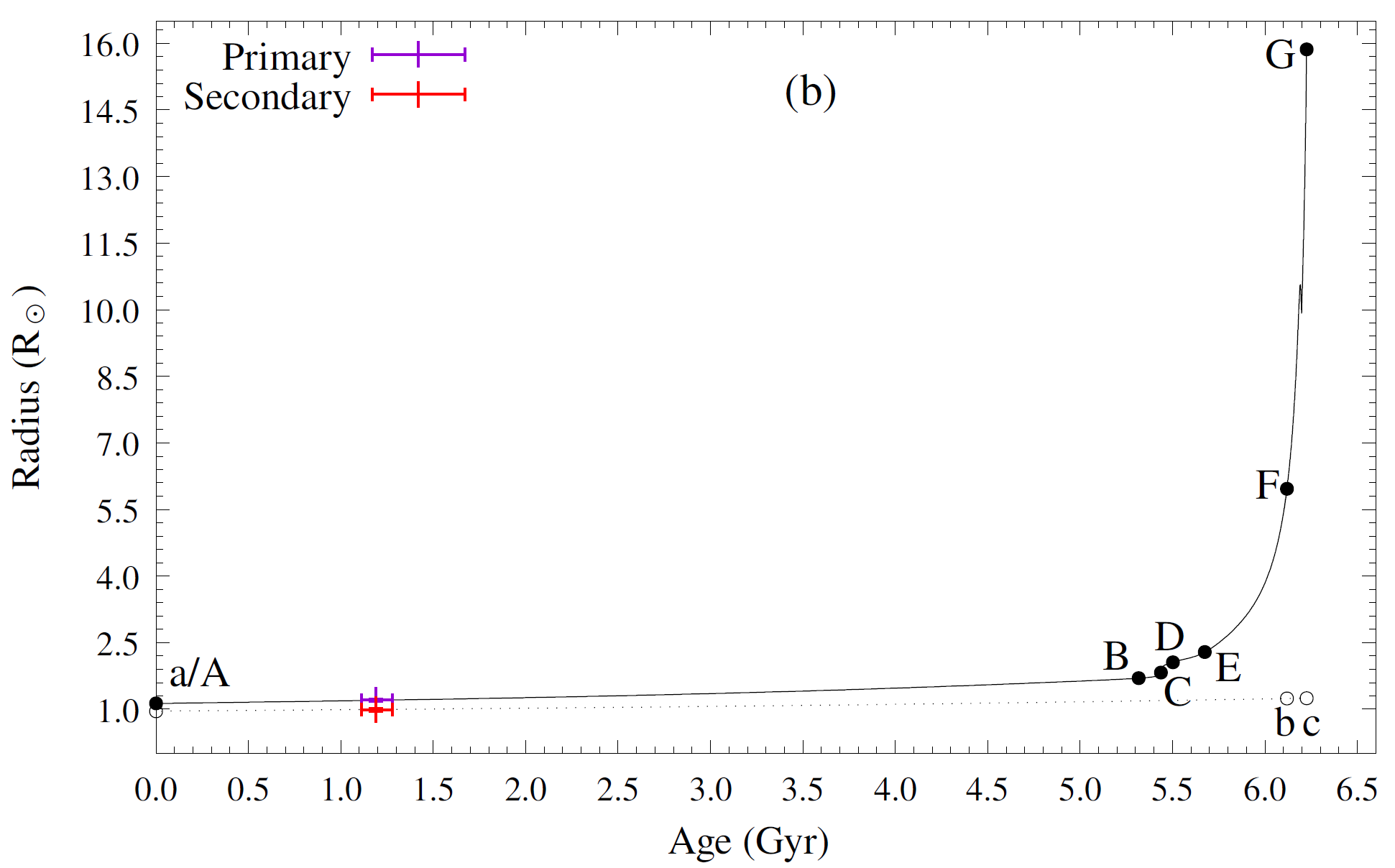}
 
     \caption{Change of orbital parameters (a), and radius of the components (b) of V454 Aur with time.}
    \label{fig:radi}
\end{figure}

\begin{figure}
	\includegraphics[width=\columnwidth]{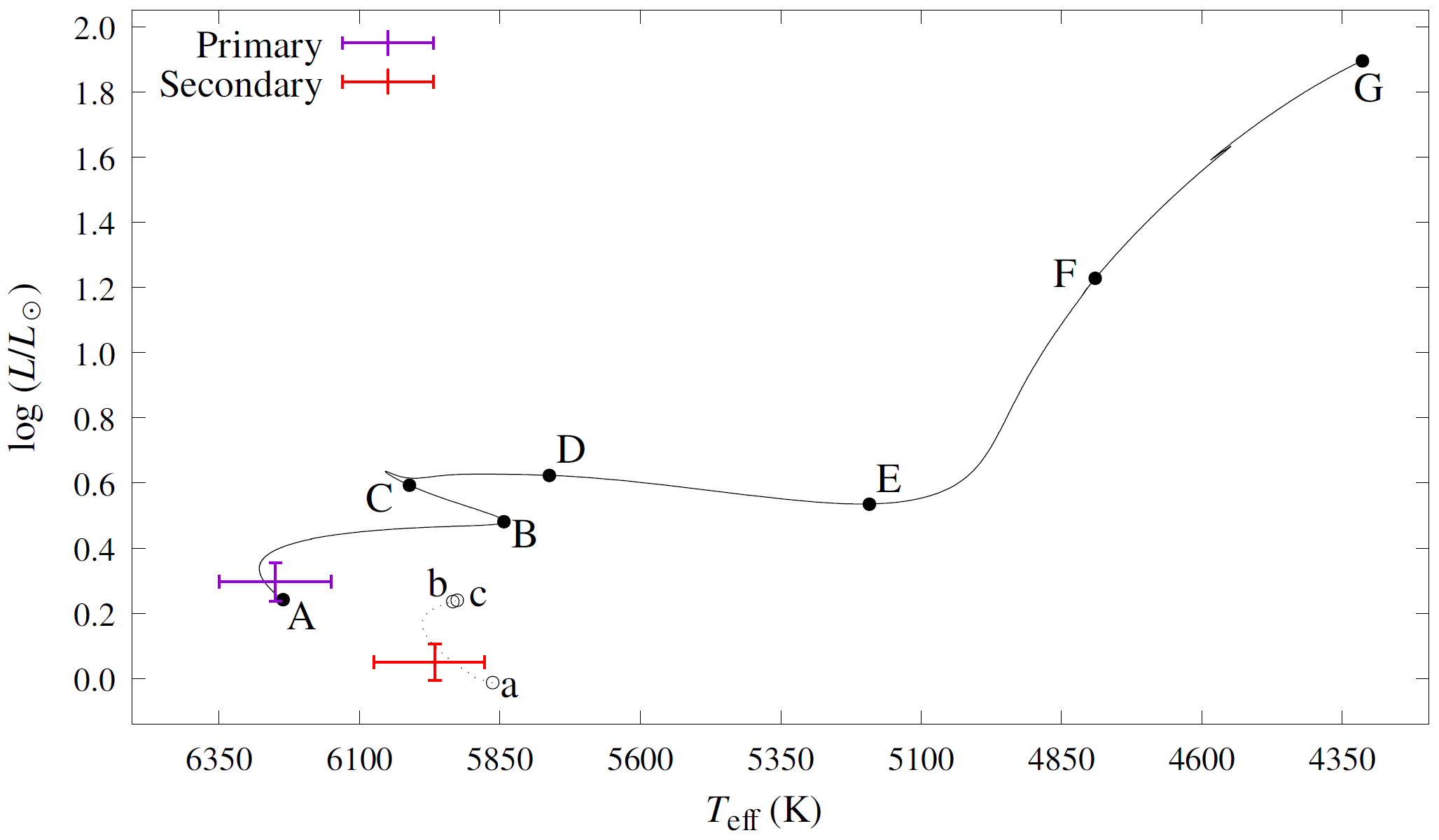}
     \caption{Evolution of V454 Aur on $T_{\rm eff}$ –$\log{L/L_\odot}$ diagram.}
    \label{fig:HR}
\end{figure}

Later on, we started binary evolution with determined initial orbital parameters and evolved the system until the primary star started to mass transfer. According to our calculations, the age of the V454 Aur is 1.19$\pm$0.09 Gyr. Taking the system as 6.23 Gyr, the mass transfer will start from the primary star, which will be on the red giant branch, to the secondary star, which still will be on the main sequence. Changes in orbital parameters and radii of the components during the evolution are presented in Figure~\ref{fig:radi}. Detailed evolution of both components of V454 Aur with timetables are given in Table~\ref{tab:v454binary} and shown in Figure~\ref{fig:HR}.

\begin{table*}
\setlength{\tabcolsep}{3pt}
\centering
    \caption{Detailed evolution of V454 Aur with time-stamps.}
    \resizebox{\textwidth}{!}
    {\begin{tabular}{lllccccccccc}
    \toprule
       \multirow{2}*{Comp} & \multirow{2}*{Mark} & \multirow{2}*{Evolutionary Status} & Age & $P$ & \multirow{2}*{$e$} & \multicolumn{3}{c}{Primary} & \multicolumn{3}{c}{Secondary} \\
       \cline{7-9}
       \cline{10-12}
        & & & (Gyr) & (day) &  & $T_\mathrm{eff}$ (K) & $\log L$ ($L_\odot$)  & $R$ ($R_\odot$) & $T_\mathrm{eff}$ (K) & $\log L$ ($L_\odot$)  & $R$ ($R_\odot$)\\
        \hline
        Pri & A & ZAMS &                 0      & 27.021 & 0.37721 & 6236 & 0.242 & 1.131 & 5863 & -0.013 & 0.955  \\
        & B & Core contraction           & 5.32 & 26.991 & 0.37628 & 5843 & 0.481 & 1.699 & 5973 & 0.208 & 1.187 \\
        & C & TAMS                       & 5.44 & 26.986 & 0.37615 & 6011 & 0.593 & 1.825 & 5969 & 0.213 & 1.194 \\
        & D & Thin H shell burning       & 5.50 & 26.978 & 0.37593 & 5762 & 0.623 & 2.056 & 5967 & 0.215 & 1.199 \\
        & E & Entering red giant phase   & 5.68 & 26.951 & 0.37519 & 5192 & 0.535 & 2.289 & 5959 & 0.221 & 1.210 \\
        & F & Circularisation of orbit   & 6.12 & 21.461 & 0       & 4790 & 1.228 & 5.968 & 5934 & 0.236 & 1.241 \\
        & G & Starting of mass transfer  & 6.23 & 21.416 & 0       & 4314 & 1.895 & 15.866& 5926 & 0.240 & 1.248 \\
        \midrule
        Sec & a & ZAMS                   & 0    & 27.021 & 0.37721 & 6236 & 0.242 & 1.131 & 5863 & -0.013 & 0.955\\
        & b & Circularisation of orbit   & 6.12 & 21.461 & 0       & 4790 & 1.228 & 5.968 & 5934 & 0.236  & 1.241\\
        & c & Starting of mass transfer  & 6.23 & 21.416 & 0       & 4314 & 1.895 & 15.866& 5926 & 0.240  & 1.248\\
        \bottomrule
    \end{tabular}
    }
    \label{tab:v454binary}
\end{table*}

\newpage
\section{Kinematics and Galactic Orbit Parameters}
With the beginning of the {\it Gaia} era, the sensitivity of astrometric measurements of stars in the Solar neighbourhood has increased. This enabled more precise determination of the kinematic and dynamic orbital parameters of nearby stars. In this study, we have focused on the analysis of V454 Aur, determining its space velocity components and Galactic orbital parameters. The proper motion components and trigonometric parallaxes of the system were obtained from the {\it Gaia} DR3 database \citep{Gaia_DR3}, and the center-of-mass velocities of the V454 Aur system, as determined in this study, are presented in Table \ref{tab:kinematic}.

The \texttt{galpy} code developed by \citet{Bovy_2015} was used to calculate the space velocity components for V454 Aur, and the uncertainties associated with these components were determined using the algorithm proposed by \citet{Johnson_1987}. These space velocity components inherently incorporate biases due to the position of stars in the Milky Way and observations from the Sun. To correct for these biases, differential rotation and local standard rest (LSR) corrections have been applied to the velocity components of the stars. Differential rotational corrections for V454 Aur were applied using the equations mentioned by \citet{Mihalas_1981}, obtaining velocity corrections of 0.04 and -0.15 km s$^{-1}$ for the $U$ and $V$ space velocity components of the system, respectively. The $W$ space velocity component, which is unaffected by differential rotation, was not corrected. For the LSR correction, the values of \citet{Coskunoglu_2011} $(U, V, W)_{\odot}=(8.83\pm 0.24, 14.19\pm 0.34, 6.57\pm 0.21)$ km s$^{-1}$ were considered and the LSR values were extracted from the space velocity components for which a differential velocity correction was applied. The relation $S_{\rm LSR}=\sqrt{U_{\rm LSR}^2+V_{\rm LSR}^2+W_{\rm LSR}^2}$ was used to calculate the total space velocity ($S_{\rm LSR}$) of the system and the finding are listed in Table \ref{tab:kinematic}. Considering the total space velocity and space velocity components of the V454 Aur, it is consistent with the value given for the young thin-disc population \citep{Leggett_1992}.

\begin{table*}[b]
\setlength{\tabcolsep}{3.5pt}
\renewcommand{\arraystretch}{1}
\small
  \centering
  \caption{Astrometric and radial velocity of V454 Aur and calculated space velocity components and Galactic orbital parameters of the system.}
\begin{tabular}{llccccccc}
\hline
\hline
\multicolumn{9}{c}{Input Parameters}\\
Star     &   $\alpha$ (J2000) & $\delta$ (J2000)  & $\mu_{\alpha}\cos\delta$ & 	$\mu_{\delta}$	 & 	$\varpi$	     &  Ref         & $V_{\rm \gamma}$  & Ref  \\
         &  (hh:mm:ss)        & (dd:mm:ss)        &       (mas yr$^{-1}$)    &   (mas yr$^{-1}$) &    (mas)          &              & (km s$^{-1}$)     &      \\
\hline
V454 Aur   & 	06:22:03.06       & 34:35:50.46	  &  -0.514$\pm$0.025       & -66.008$\pm$0.019 &  15.367$\pm$0.022 &   [1]        & -40.48$\pm$0.10    & [2]  \\	
\hline
\multicolumn{9}{c}{Output Parameters}\\
\hline
\hline
Star  &  $U_{\rm LSR}$     & $V_{\rm LSR}$  & $W_{\rm LSR}$  & $S_{\rm LSR}$	& $R_{\rm a}$	 & $R_{\rm p}$       & $Z_{\rm max}$ & $e$  \\
      &  (km s$^{-1}$)     & (km s$^{-1}$)  & (km s$^{-1}$)  & (km s$^{-1}$)    & (pc)           & (pc)              & (pc)          &      \\
\hline
V454 Aur   & 46.80$\pm$0.10  & -4.61$\pm$0.03	& -9.22$\pm$0.02 & 47.92$\pm$0.11   & 10038$\pm$40   & 7044$\pm$20        & 370$\pm$1    & 0.175$\pm$0.001\\
\hline
    \end{tabular}
    \\
Ref: [1] \citet{Gaia_DR3}, [2] This study 	
    \label{tab:kinematic}
\end{table*} 

\begin{figure}[h!]
\centering\includegraphics[width=1\linewidth]{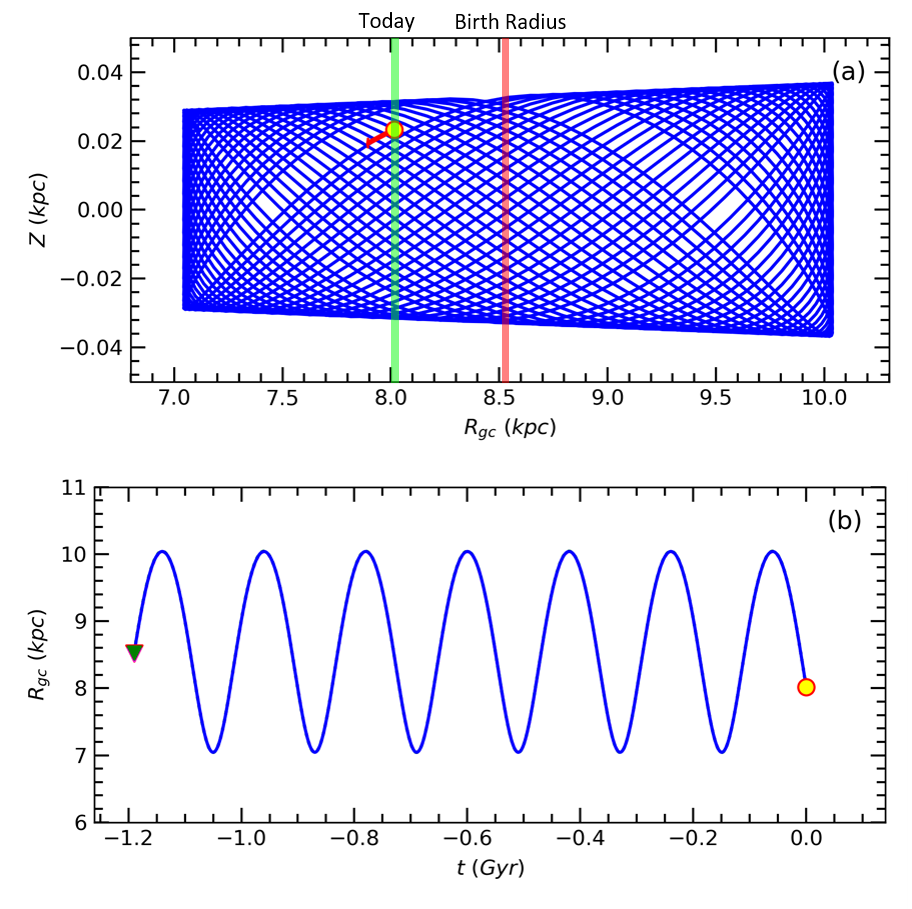}
\caption{The Galactic orbits and birth radii of V454 Aur in the $Z \times R_{\rm gc}$ (a) and $R_{\rm gc} \times t$ (b) diagrams. The filled yellow circles and triangles show the today and birth positions, respectively. The red arrow is the motion vector of V454 Aur in today. The green and pink dotted lines show the orbit when errors in input parameters are considered, while the green and pink filled triangles represent the birth locations of the V454 Aur based on the lower and upper error estimates.}
\label{fig:galactic_orbits}
\end{figure}

The \texttt{galpy} code \citep{Bovy_2015} was also used to compute the Galactic orbital parameters of the V454 Aur. For the Galactic potentials needed for the Galactic orbit calculations, \texttt{MWPotential2014} was used, which was created specifically for the Galaxy. For the system to form closed orbits around the Galactic centre, a timescale of 3 Gyr in 2 Myr steps was used. The Galactic orbital calculations resulted in the determination of several parameters, including the perigalactic distance ($R_{\rm p}$), apogalactic distance ($R_{\rm a}$), maximum distance from the Galactic plane ($Z_{\rm max}$), and eccentricity ($e$) of the Galactic orbits. These parameters are also listed in Table \ref{tab:kinematic}. The positions of the system according to their distances from the Galactic center ($R_{\rm gc}$) and perpendicular to the Galactic plane ($Z$) at different times are shown in Figure \ref{fig:galactic_orbits}a. The \texttt{galpy} results show that the V454 Aur has a slightly flattened Galactic orbit. Moreover, the fact that the system is at $Z=11$ pc ($Z=d\times \sin b$) from the Galactic plane is an indication that V454 Aur may belong to the young thin-disc component of Milk Way \citep{Guctekin_2019}.

The Galactic orbits for the V454 Aur system on $Z \times R_{\rm gc}$ and $R_{\rm gc} \times t$ diagrams are shown in Figure \ref{fig:galactic_orbits}. The panels in Figure \ref{fig:galactic_orbits} show side views of the V454 Aur motions as functions of distance from the Galactic center and the Galactic plane, respectively \citep{Tasdemir_2023}. In Figure \ref{fig:galactic_orbits}b the birth and present-day positions for the V454 Aur are shown with yellow-filled triangle and circle, respectively \citep{Yontan_2022}. The eccentricity of the orbit of the V454 Aur does not exceed the value of 0.18. The distances from the Galactic plane reach out maximum at $Z_{\rm max}=370\pm1$ pc for V454 Aur. These results show that the V454 Aur belongs to the young thin disc of the Milky Way. The birthplace of the system was also investigated by running the binary system age ($t=1.19\pm0.08$ Gyr) calculated in this work back in time using the \texttt{galpy} program \citep{Yontan-Canbay_2023}. The birth radius of the binary system was determined as $R_{\rm Birth}=8.52\pm0.02$ kpc. These findings represent that the V454 Aur was born almost in the solar-abundance region around the Solar circle.

\section{Conclusions}

Eclipsing binary systems are the foundation of the studies of stellar astrophysics as they provide accurate stellar parameters, which can be used in any area of astrophysics. In this study, we have derived, for the first time, the fundamental parameters of V454 Aur (HD 44192) including temperature, metallicity, and age by combining photometric radial velocities, precise photometric data, and SED data. We have calculated the mass of the components in V454 Aur better than 1.5\% as 1.173 M$_\odot$ and 1.045 M$_\odot$ for the primary and the secondary component, respectively, and the radii of the components better than 3\% as 1.203 R$_\odot$ and 0.993 R$_\odot$ for the primary and the secondary component, respectively. Our distance calculation is in excellent agreement with the \textit{Gaia} DR3 one \citep{Gaia_DR3}, which is in the 2\% error range if the errors are included, which indicates that our calculations are accurate. Mass-luminosity parameters that we have derived for the components are slight smaller than \cite{Eker2018}, which is expected considering the position of the V454 Aur on the main sequence. According to our evolution analysis, the components of V454 Aur are still in the main sequence and a little richer than solar metallicity, $z=0.017$, and the system is long away from the mass transfer. The initial orbital parameters of the system have been derived by using state-of-the-art evolutionary code \texttt{MESA} and evolutionary status in several phases were noted in Table~\ref{tab:v454binary}. According to our calculations, the age of the system is $1.19_{-0.09}^{\,+0.08}$ Gyr, and it will start to mass transfer between components in 5 Gyr when the primary component is in the red giant branch and the secondary component is still on main-sequence. The calculations of detailed evolutionary steps for eclipsing binaries are important because these calculations could shed light on, generally, understanding the properties of current semi-detached binaries. 

Considering the dynamical orbital parameters and the age of the V454 Aur system, it was determined that it formed in a region around the Solar circle. This result is also consistent with the metal abundance assumed for the V454 Aur. There are still very few systems where their evolutionary phases and birthplace have been revealed. We believe that finding the initial orbital properties of eclipsing binaries and the location of their birthplaces would help to understand the formation mechanism of eclipsing binaries in detail.

\section*{Acknowledgements}
We are grateful to the anonymous referees for their valuable suggestions. This paper includes data collected with the \textit{TESS} mission, obtained from the MAST data archive at the Space Telescope Science Institute (STScI). Funding for the \textit{TESS} mission is provided by the NASA Explorer Program. STScI is operated by the Association of Universities for Research in Astronomy, Inc., under NASA contract NAS 5–26555. This research has made use of the SIMBAD database, operated at CDS, Strasbourg, France. This work has made use of data from the European Space Agency (ESA) mission \textit{Gaia} (\url{https://www.cosmos.esa.int/gaia}), processed by the \textit{Gaia} Data Processing and Analysis Consortium (DPAC, \url{https://www.cosmos.esa.int/web/gaia/dpac/consortium}). Funding for the DPAC has been provided by national institutions, in particular, the institutions participating in the \textit{Gaia} Multilateral Agreement. This research has made use of NASA’s Astrophysics Data System. 

In addition to the Python packages referenced in \S3 and \S4, this research has made use of the following packages:
\texttt{Astropy} \citep{astro1,astro2,astro3}, \texttt{corner} \citep{corner}, \texttt{Matplotlib} \citep{matplotlib}, \texttt{NumPy} \citep{numpy}, and \texttt{SciPy} \citep{scipy}.

\bibliographystyle{mnras}
\bibliography{reference} 

\begin{thebibliography}{}
\makeatletter
\relax
\def\mn@urlcharsother{\let\do\@makeother \do\$\do\&\do\#\do\^\do\_\do\%\do\~}
\def\mn@doi{\begingroup\mn@urlcharsother \@ifnextchar [ {\mn@doi@} {\mn@doi@[]}}
\def\mn@doi@[#1]#2{\def\@tempa{#1}\ifx\@tempa\@empty \href {http://dx.doi.org/#2} {doi:#2}\else \href {http://dx.doi.org/#2} {#1}\fi \endgroup}
\def\mn@eprint#1#2{\mn@eprint@#1:#2::\@nil}
\def\mn@eprint@arXiv#1{\href {http://arxiv.org/abs/#1} {{\tt arXiv:#1}}}
\def\mn@eprint@dblp#1{\href {http://dblp.uni-trier.de/rec/bibtex/#1.xml} {dblp:#1}}
\def\mn@eprint@#1:#2:#3:#4\@nil{\def\@tempa {#1}\def\@tempb {#2}\def\@tempc {#3}\ifx \@tempc \@empty \let \@tempc \@tempb \let \@tempb \@tempa \fi \ifx \@tempb \@empty \def\@tempb {arXiv}\fi \@ifundefined {mn@eprint@\@tempb}{\@tempb:\@tempc}{\expandafter \expandafter \csname mn@eprint@\@tempb\endcsname \expandafter{\@tempc}}}

\bibitem[\protect\citeauthoryear{{Angulo} et~al.,}{{Angulo} et~al.}{1999}]{Angulo1999}
{Angulo} C.,  et~al., 1999, \mn@doi [\nphysa] {10.1016/S0375-9474(99)00030-5}, \href {https://ui.adsabs.harvard.edu/abs/1999NuPhA.656....3A} {656, 3}

\bibitem[\protect\citeauthoryear{{Astropy Collaboration} et~al.,}{{Astropy Collaboration} et~al.}{2013}]{astro1}
{Astropy Collaboration} et~al., 2013, \mn@doi [\aap] {10.1051/0004-6361/201322068}, \href {https://ui.adsabs.harvard.edu/abs/2013A&A...558A..33A} {558, A33}

\bibitem[\protect\citeauthoryear{{Astropy Collaboration} et~al.,}{{Astropy Collaboration} et~al.}{2018}]{astro2}
{Astropy Collaboration} et~al., 2018, \mn@doi [\aj] {10.3847/1538-3881/aabc4f}, \href {https://ui.adsabs.harvard.edu/abs/2018AJ....156..123A} {156, 123}

\bibitem[\protect\citeauthoryear{{Astropy Collaboration} et~al.,}{{Astropy Collaboration} et~al.}{2022}]{astro3}
{Astropy Collaboration} et~al., 2022, \mn@doi [\apj] {10.3847/1538-4357/ac7c74}, \href {https://ui.adsabs.harvard.edu/abs/2022ApJ...935..167A} {935, 167}

\bibitem[\protect\citeauthoryear{{Bak{\i}{\c{s}}} \& {Eker}}{{Bak{\i}{\c{s}}} \& {Eker}}{2022}]{Bakis2022}
{Bak{\i}{\c{s}}} V.,  {Eker} Z.,  2022, \mn@doi [\actaa] {10.32023/0001-5237/72.3.4}, \href {https://ui.adsabs.harvard.edu/abs/2022AcA....72..195B} {72, 195}

\bibitem[\protect\citeauthoryear{{Baranne}, {Mayor}  \& {Poncet}}{{Baranne} et~al.}{1979}]{Baranne}
{Baranne} A.,  {Mayor} M.,   {Poncet} J.~L.,  1979, \mn@doi [Vistas in Astronomy] {10.1016/0083-6656(79)90016-3}, \href {https://ui.adsabs.harvard.edu/abs/1979VA.....23..279B} {23, 279}

\bibitem[\protect\citeauthoryear{{Benedict} et~al.,}{{Benedict} et~al.}{2016}]{Benedict}
{Benedict} G.~F.,  et~al., 2016, \mn@doi [\aj] {10.3847/0004-6256/152/5/141}, \href {https://ui.adsabs.harvard.edu/abs/2016AJ....152..141B} {152, 141}

\bibitem[\protect\citeauthoryear{{Blouin}, {Shaffer}, {Saumon}  \& {Starrett}}{{Blouin} et~al.}{2020}]{Blouin2020}
{Blouin} S.,  {Shaffer} N.~R.,  {Saumon} D.,   {Starrett} C.~E.,  2020, \mn@doi [\apj] {10.3847/1538-4357/ab9e75}, \href {https://ui.adsabs.harvard.edu/abs/2020ApJ...899...46B} {899, 46}

\bibitem[\protect\citeauthoryear{{Bovy}}{{Bovy}}{2015}]{Bovy_2015}
{Bovy} J.,  2015, \mn@doi [\apjs] {10.1088/0067-0049/216/2/29}, \href {https://ui.adsabs.harvard.edu/abs/2015ApJS..216...29B} {216, 29}

\bibitem[\protect\citeauthoryear{{Casagrande}, {Sch{\"o}nrich}, {Asplund}, {Cassisi}, {Ram{\'\i}rez}, {Mel{\'e}ndez}, {Bensby}  \& {Feltzing}}{{Casagrande} et~al.}{2011}]{Casagrande}
{Casagrande} L.,  {Sch{\"o}nrich} R.,  {Asplund} M.,  {Cassisi} S.,  {Ram{\'\i}rez} I.,  {Mel{\'e}ndez} J.,  {Bensby} T.,   {Feltzing} S.,  2011, \mn@doi [\aap] {10.1051/0004-6361/201016276}, \href {https://ui.adsabs.harvard.edu/abs/2011A&A...530A.138C} {530, A138}

\bibitem[\protect\citeauthoryear{{Cassisi}, {Potekhin}, {Pietrinferni}, {Catelan}  \& {Salaris}}{{Cassisi} et~al.}{2007}]{Cassisi2007}
{Cassisi} S.,  {Potekhin} A.~Y.,  {Pietrinferni} A.,  {Catelan} M.,   {Salaris} M.,  2007, \mn@doi [\apj] {10.1086/516819}, \href {https://ui.adsabs.harvard.edu/abs/2007ApJ...661.1094C} {661, 1094}

\bibitem[\protect\citeauthoryear{{Chabrier}}{{Chabrier}}{2003}]{Chabrier}
{Chabrier} G.,  2003, \mn@doi [\pasp] {10.1086/376392}, \href {https://ui.adsabs.harvard.edu/abs/2003PASP..115..763C} {115, 763}

\bibitem[\protect\citeauthoryear{{Chugunov}, {Dewitt}  \& {Yakovlev}}{{Chugunov} et~al.}{2007}]{Chugunov2007}
{Chugunov} A.~I.,  {Dewitt} H.~E.,   {Yakovlev} D.~G.,  2007, \mn@doi [\prd] {10.1103/PhysRevD.76.025028}, \href {https://ui.adsabs.harvard.edu/abs/2007PhRvD..76b5028C} {76, 025028}

\bibitem[\protect\citeauthoryear{Co{\c{s}}kuno{\v{g}}lu et~al.,}{Co{\c{s}}kuno{\v{g}}lu et~al.}{2011}]{Coskunoglu_2011}
Co{\c{s}}kuno{\v{g}}lu B.,  et~al., 2011, \mn@doi [\mnras] {10.1111/j.1365-2966.2010.17983.x}, \href {https://ui.adsabs.harvard.edu/abs/2011MNRAS.412.1237C} {412, 1237}

\bibitem[\protect\citeauthoryear{{Cyburt} et~al.,}{{Cyburt} et~al.}{2010}]{Cyburt2010}
{Cyburt} R.~H.,  et~al., 2010, \mn@doi [\apjs] {10.1088/0067-0049/189/1/240}, \href {https://ui.adsabs.harvard.edu/abs/2010ApJS..189..240C} {189, 240}

\bibitem[\protect\citeauthoryear{{Eggleton}}{{Eggleton}}{1983}]{Eggleton1983}
{Eggleton} P.~P.,  1983, \mn@doi [\apj] {10.1086/160960}, \href {https://ui.adsabs.harvard.edu/abs/1983ApJ...268..368E} {268, 368}

\bibitem[\protect\citeauthoryear{{Eker} \& {Bak{\i}{\c{s}}}}{{Eker} \& {Bak{\i}{\c{s}}}}{2023}]{Eker2023}
{Eker} Z.,  {Bak{\i}{\c{s}}} V.,  2023, \mn@doi [\mnras] {10.1093/mnras/stad1563}, \href {https://ui.adsabs.harvard.edu/abs/2023MNRAS.523.2440E} {523, 2440}

\bibitem[\protect\citeauthoryear{{Eker} et~al.,}{{Eker} et~al.}{2015}]{Eker2015}
{Eker} Z.,  et~al., 2015, \mn@doi [\aj] {10.1088/0004-6256/149/4/131}, \href {https://ui.adsabs.harvard.edu/abs/2015AJ....149..131E} {149, 131}

\bibitem[\protect\citeauthoryear{{Eker} et~al.,}{{Eker} et~al.}{2018}]{Eker2018}
{Eker} Z.,  et~al., 2018, \mn@doi [\mnras] {10.1093/mnras/sty1834}, \href {https://ui.adsabs.harvard.edu/abs/2018MNRAS.479.5491E} {479, 5491}

\bibitem[\protect\citeauthoryear{{Eker}, {Soydugan}  \& {Bilir}}{{Eker} et~al.}{2024}]{Eker2024}
{Eker} Z.,  {Soydugan} F.,   {Bilir} S.,  2024, \mn@doi [arXiv e-prints] {10.48550/arXiv.2402.07947}, \href {https://ui.adsabs.harvard.edu/abs/2024arXiv240207947E} {p. arXiv:2402.07947}

\bibitem[\protect\citeauthoryear{{Ferguson}, {Alexander}, {Allard}, {Barman}, {Bodnarik}, {Hauschildt}, {Heffner-Wong}  \& {Tamanai}}{{Ferguson} et~al.}{2005}]{Ferguson2005}
{Ferguson} J.~W.,  {Alexander} D.~R.,  {Allard} F.,  {Barman} T.,  {Bodnarik} J.~G.,  {Hauschildt} P.~H.,  {Heffner-Wong} A.,   {Tamanai} A.,  2005, \mn@doi [\apj] {10.1086/428642}, \href {https://ui.adsabs.harvard.edu/abs/2005ApJ...623..585F} {623, 585}

\bibitem[\protect\citeauthoryear{{Foreman-Mackey}}{{Foreman-Mackey}}{2016}]{corner}
{Foreman-Mackey} D.,  2016, \mn@doi [The Journal of Open Source Software] {10.21105/joss.00024}, \href {https://ui.adsabs.harvard.edu/abs/2016JOSS....1...24F} {1, 24}

\bibitem[\protect\citeauthoryear{{Foreman-Mackey}, {Hogg}, {Lang}  \& {Goodman}}{{Foreman-Mackey} et~al.}{2013}]{mcmc}
{Foreman-Mackey} D.,  {Hogg} D.~W.,  {Lang} D.,   {Goodman} J.,  2013, \mn@doi [\pasp] {10.1086/670067}, \href {https://ui.adsabs.harvard.edu/abs/2013PASP..125..306F} {125, 306}

\bibitem[\protect\citeauthoryear{{Fuller}, {Fowler}  \& {Newman}}{{Fuller} et~al.}{1985}]{Fuller1985}
{Fuller} G.~M.,  {Fowler} W.~A.,   {Newman} M.~J.,  1985, \mn@doi [\apj] {10.1086/163208}, \href {https://ui.adsabs.harvard.edu/abs/1985ApJ...293....1F} {293, 1}

\bibitem[\protect\citeauthoryear{{Gaia Collaboration} et~al.,}{{Gaia Collaboration} et~al.}{2023}]{Gaia_DR3}
{Gaia Collaboration} et~al., 2023, \mn@doi [\aap] {10.1051/0004-6361/202243940}, \href {https://ui.adsabs.harvard.edu/abs/2021A&A...649A...1G} {674, A1}

\bibitem[\protect\citeauthoryear{{Griffin}}{{Griffin}}{2001}]{Griffin2001}
{Griffin} R.~F.,  2001, The Observatory, \href {https://ui.adsabs.harvard.edu/abs/2001Obs...121..315G} {121, 315}

\bibitem[\protect\citeauthoryear{{Harris} et~al.,}{{Harris} et~al.}{2020}]{numpy}
{Harris} C.~R.,  et~al., 2020, \mn@doi [\nat] {10.1038/s41586-020-2649-2}, \href {https://ui.adsabs.harvard.edu/abs/2020Natur.585..357H} {585, 357}

\bibitem[\protect\citeauthoryear{{Holmberg}, {Nordstr{\"o}m}  \& {Andersen}}{{Holmberg} et~al.}{2009}]{Holmberg}
{Holmberg} J.,  {Nordstr{\"o}m} B.,   {Andersen} J.,  2009, \mn@doi [\aap] {10.1051/0004-6361/200811191}, \href {https://ui.adsabs.harvard.edu/abs/2009A&A...501..941H} {501, 941}

\bibitem[\protect\citeauthoryear{{Hunter}}{{Hunter}}{2007}]{matplotlib}
{Hunter} J.~D.,  2007, \mn@doi [Computing in Science and Engineering] {10.1109/MCSE.2007.55}, \href {https://ui.adsabs.harvard.edu/abs/2007CSE.....9...90H} {9, 90}

\bibitem[\protect\citeauthoryear{{Hurley}, {Tout}  \& {Pols}}{{Hurley} et~al.}{2002}]{Hurley}
{Hurley} J.~R.,  {Tout} C.~A.,   {Pols} O.~R.,  2002, \mn@doi [\mnras] {10.1046/j.1365-8711.2002.05038.x}, \href {https://ui.adsabs.harvard.edu/abs/2002MNRAS.329..897H} {329, 897}

\bibitem[\protect\citeauthoryear{{Iglesias} \& {Rogers}}{{Iglesias} \& {Rogers}}{1993}]{Iglesias1993}
{Iglesias} C.~A.,  {Rogers} F.~J.,  1993, \mn@doi [\apj] {10.1086/172958}, \href {https://ui.adsabs.harvard.edu/abs/1993ApJ...412..752I} {412, 752}

\bibitem[\protect\citeauthoryear{{Iglesias} \& {Rogers}}{{Iglesias} \& {Rogers}}{1996}]{Iglesias1996}
{Iglesias} C.~A.,  {Rogers} F.~J.,  1996, \mn@doi [\apj] {10.1086/177381}, \href {https://ui.adsabs.harvard.edu/abs/1996ApJ...464..943I} {464, 943}

\bibitem[\protect\citeauthoryear{{Irwin}}{{Irwin}}{2004}]{Irwin2004}
{Irwin} A.~W.,  2004, The FreeEOS Code for Calculating the Equation of State for Stellar Interiors, \url {http://freeeos.sourceforge.net/}

\bibitem[\protect\citeauthoryear{{Itoh}, {Hayashi}, {Nishikawa}  \& {Kohyama}}{{Itoh} et~al.}{1996}]{Itoh1996}
{Itoh} N.,  {Hayashi} H.,  {Nishikawa} A.,   {Kohyama} Y.,  1996, \mn@doi [\apjs] {10.1086/192264}, \href {https://ui.adsabs.harvard.edu/abs/1996ApJS..102..411I} {102, 411}

\bibitem[\protect\citeauthoryear{{Jermyn}, {Schwab}, {Bauer}, {Timmes}  \& {Potekhin}}{{Jermyn} et~al.}{2021}]{Jermyn2021}
{Jermyn} A.~S.,  {Schwab} J.,  {Bauer} E.,  {Timmes} F.~X.,   {Potekhin} A.~Y.,  2021, \mn@doi [\apj] {10.3847/1538-4357/abf48e}, \href {https://ui.adsabs.harvard.edu/abs/2021ApJ...913...72J} {913, 72}

\bibitem[\protect\citeauthoryear{{Jermyn} et~al.,}{{Jermyn} et~al.}{2023}]{Jermyn2023}
{Jermyn} A.~S.,  et~al., 2023, \mn@doi [\apjs] {10.3847/1538-4365/acae8d}, \href {https://ui.adsabs.harvard.edu/abs/2023ApJS..265...15J} {265, 15}

\bibitem[\protect\citeauthoryear{{Johnson} \& {Soderblom}}{{Johnson} \& {Soderblom}}{1987}]{Johnson_1987}
{Johnson} D. R.~H.,  {Soderblom} D.~R.,  1987, \mn@doi [\aj] {10.1086/114370}, \href {https://ui.adsabs.harvard.edu/abs/1987AJ.....93..864J} {93, 864}

\bibitem[\protect\citeauthoryear{{Kwee} \& {van Woerden}}{{Kwee} \& {van Woerden}}{1956}]{Kwee}
{Kwee} K.~K.,  {van Woerden} H.,  1956, \bain, \href {https://ui.adsabs.harvard.edu/abs/1956BAN....12..327K} {12, 327}

\bibitem[\protect\citeauthoryear{{Langanke} \& {Mart{\'{\i}}nez-Pinedo}}{{Langanke} \& {Mart{\'{\i}}nez-Pinedo}}{2000}]{Langanke2000}
{Langanke} K.,  {Mart{\'{\i}}nez-Pinedo} G.,  2000, \mn@doi [Nuclear Physics A] {10.1016/S0375-9474(00)00131-7}, \href {https://ui.adsabs.harvard.edu/abs/2000NuPhA.673..481L} {673, 481}

\bibitem[\protect\citeauthoryear{Leggett}{Leggett}{1992}]{Leggett_1992}
Leggett S.~K.,  1992, \mn@doi [\apjs] {10.1086/191720}, \href {https://ui.adsabs.harvard.edu/abs/1992ApJS...82..351L} {82, 351}

\bibitem[\protect\citeauthoryear{{Lightkurve Collaboration} et~al.,}{{Lightkurve Collaboration} et~al.}{2018}]{lk}
{Lightkurve Collaboration} et~al., 2018, {Lightkurve: Kepler and TESS time series analysis in Python}, Astrophysics Source Code Library (\mn@eprint {ascl} {1812.013})

\bibitem[\protect\citeauthoryear{{Mihalas} \& {Binney}}{{Mihalas} \& {Binney}}{1981}]{Mihalas_1981}
{Mihalas} D.,  {Binney} J.,  1981, {Galactic astronomy. Structure and kinematics}

\bibitem[\protect\citeauthoryear{{Moe} \& {Di Stefano}}{{Moe} \& {Di Stefano}}{2017}]{Moe}
{Moe} M.,  {Di Stefano} R.,  2017, \mn@doi [\apjs] {10.3847/1538-4365/aa6fb6}, \href {https://ui.adsabs.harvard.edu/abs/2017ApJS..230...15M} {230, 15}

\bibitem[\protect\citeauthoryear{{Nordstr{\"o}m} et~al.,}{{Nordstr{\"o}m} et~al.}{2004}]{Nordstrom}
{Nordstr{\"o}m} B.,  et~al., 2004, \mn@doi [\aap] {10.1051/0004-6361:20035959}, \href {https://ui.adsabs.harvard.edu/abs/2004A&A...418..989N} {418, 989}

\bibitem[\protect\citeauthoryear{{Oda}, {Hino}, {Muto}, {Takahara}  \& {Sato}}{{Oda} et~al.}{1994}]{Oda1994}
{Oda} T.,  {Hino} M.,  {Muto} K.,  {Takahara} M.,   {Sato} K.,  1994, \mn@doi [Atomic Data and Nuclear Data Tables] {10.1006/adnd.1994.1007}, \href {https://ui.adsabs.harvard.edu/abs/1994ADNDT..56..231O} {56, 231}

\bibitem[\protect\citeauthoryear{{Paegert}, {Stassun}, {Collins}, {Pepper}, {Torres}, {Jenkins}, {Twicken}  \& {Latham}}{{Paegert} et~al.}{2022}]{Paegert2022}
{Paegert} M.,  {Stassun} K.~G.,  {Collins} K.~A.,  {Pepper} J.,  {Torres} G.,  {Jenkins} J.,  {Twicken} J.~D.,   {Latham} D.~W.,  2022, {VizieR Online Data Catalog: TESS Input Catalog version 8.2 (TIC v8.2) (Paegert+, 2021)}, VizieR On-line Data Catalog: IV/39. Originally published in: 2021arXiv210804778P

\bibitem[\protect\citeauthoryear{{Paxton}, {Bildsten}, {Dotter}, {Herwig}, {Lesaffre}  \& {Timmes}}{{Paxton} et~al.}{2011}]{Paxton2011}
{Paxton} B.,  {Bildsten} L.,  {Dotter} A.,  {Herwig} F.,  {Lesaffre} P.,   {Timmes} F.,  2011, \mn@doi [\apjs] {10.1088/0067-0049/192/1/3}, \href {https://ui.adsabs.harvard.edu/abs/2011ApJS..192....3P} {192, 3}

\bibitem[\protect\citeauthoryear{{Paxton} et~al.,}{{Paxton} et~al.}{2013}]{Paxton2013}
{Paxton} B.,  et~al., 2013, \mn@doi [\apjs] {10.1088/0067-0049/208/1/4}, \href {https://ui.adsabs.harvard.edu/abs/2013ApJS..208....4P} {208, 4}

\bibitem[\protect\citeauthoryear{{Paxton} et~al.,}{{Paxton} et~al.}{2015}]{Paxton2015}
{Paxton} B.,  et~al., 2015, \mn@doi [\apjs] {10.1088/0067-0049/220/1/15}, \href {https://ui.adsabs.harvard.edu/abs/2015ApJS..220...15P} {220, 15}

\bibitem[\protect\citeauthoryear{{Paxton} et~al.,}{{Paxton} et~al.}{2018}]{Paxton2018}
{Paxton} B.,  et~al., 2018, \mn@doi [\apjs] {10.3847/1538-4365/aaa5a8}, \href {https://ui.adsabs.harvard.edu/abs/2018ApJS..234...34P} {234, 34}

\bibitem[\protect\citeauthoryear{{Paxton} et~al.,}{{Paxton} et~al.}{2019}]{Paxton2019}
{Paxton} B.,  et~al., 2019, \mn@doi [\apjs] {10.3847/1538-4365/ab2241}, \href {https://ui.adsabs.harvard.edu/abs/2019ApJS..243...10P} {243, 10}

\bibitem[\protect\citeauthoryear{{Perryman} et~al.,}{{Perryman} et~al.}{1997}]{Perryman}
{Perryman} M.~A.~C.,  et~al., 1997, \aap, \href {https://ui.adsabs.harvard.edu/abs/1997A&A...323L..49P} {323, L49}

\bibitem[\protect\citeauthoryear{{Potekhin} \& {Chabrier}}{{Potekhin} \& {Chabrier}}{2010}]{Potekhin2010}
{Potekhin} A.~Y.,  {Chabrier} G.,  2010, \mn@doi [Contributions to Plasma Physics] {10.1002/ctpp.201010017}, \href {https://ui.adsabs.harvard.edu/abs/2010CoPP...50...82P} {50, 82}

\bibitem[\protect\citeauthoryear{{Poutanen}}{{Poutanen}}{2017}]{Poutanen2017}
{Poutanen} J.,  2017, \mn@doi [\apj] {10.3847/1538-4357/835/2/119}, \href {https://ui.adsabs.harvard.edu/abs/2017ApJ...835..119P} {835, 119}

\bibitem[\protect\citeauthoryear{{Pr{\v{s}}a}}{{Pr{\v{s}}a}}{2020}]{Prsa2020}
{Pr{\v{s}}a} A.,  2020, \mn@doi [Contributions of the Astronomical Observatory Skalnate Pleso] {10.31577/caosp.2020.50.2.565}, \href {https://ui.adsabs.harvard.edu/abs/2020CoSka..50..565P} {50, 565}

\bibitem[\protect\citeauthoryear{{Pr{\v{s}}a} \& {Zwitter}}{{Pr{\v{s}}a} \& {Zwitter}}{2005}]{phoebe}
{Pr{\v{s}}a} A.,  {Zwitter} T.,  2005, \mn@doi [\apj] {10.1086/430591}, \href {https://ui.adsabs.harvard.edu/abs/2005ApJ...628..426P} {628, 426}

\bibitem[\protect\citeauthoryear{{Pr{\v{s}}a} et~al.,}{{Pr{\v{s}}a} et~al.}{2022}]{PrsaTESS}
{Pr{\v{s}}a} A.,  et~al., 2022, \mn@doi [\apjs] {10.3847/1538-4365/ac324a}, \href {https://ui.adsabs.harvard.edu/abs/2022ApJS..258...16P} {258, 16}

\bibitem[\protect\citeauthoryear{{Rappaport}, {Verbunt}  \& {Joss}}{{Rappaport} et~al.}{1983}]{magnetic}
{Rappaport} S.,  {Verbunt} F.,   {Joss} P.~C.,  1983, \mn@doi [\apj] {10.1086/161569}, \href {https://ui.adsabs.harvard.edu/abs/1983ApJ...275..713R} {275, 713}

\bibitem[\protect\citeauthoryear{{Ricker} et~al.,}{{Ricker} et~al.}{2015}]{Ricker}
{Ricker} G.~R.,  et~al., 2015, \mn@doi [Journal of Astronomical Telescopes, Instruments, and Systems] {10.1117/1.JATIS.1.1.014003}, \href {https://ui.adsabs.harvard.edu/abs/2015JATIS...1a4003R} {1, 014003}

\bibitem[\protect\citeauthoryear{{Ritter}}{{Ritter}}{1988}]{Ritter1988}
{Ritter} H.,  1988, \aap, \href {https://ui.adsabs.harvard.edu/abs/1988A%26A...202...93R} {202, 93}

\bibitem[\protect\citeauthoryear{{Rogers} \& {Nayfonov}}{{Rogers} \& {Nayfonov}}{2002}]{Rogers2002}
{Rogers} F.~J.,  {Nayfonov} A.,  2002, \mn@doi [\apj] {10.1086/341894}, \href {https://ui.adsabs.harvard.edu/abs/2002ApJ...576.1064R} {576, 1064}

\bibitem[\protect\citeauthoryear{{Rosales}, {Mennickent}, {Schleicher}  \& {Senhadji}}{{Rosales} et~al.}{2019}]{Rosales}
{Rosales} J.~A.,  {Mennickent} R.~E.,  {Schleicher} D.~R.~G.,   {Senhadji} A.~A.,  2019, \mn@doi [\mnras] {10.1093/mnras/sty3117}, \href {https://ui.adsabs.harvard.edu/abs/2019MNRAS.483..862R} {483, 862}

\bibitem[\protect\citeauthoryear{{Saumon}, {Chabrier}  \& {van Horn}}{{Saumon} et~al.}{1995}]{Saumon1995}
{Saumon} D.,  {Chabrier} G.,   {van Horn} H.~M.,  1995, \mn@doi [\apjs] {10.1086/192204}, \href {https://ui.adsabs.harvard.edu/abs/1995ApJS...99..713S} {99, 713}

\bibitem[\protect\citeauthoryear{{Serenelli} et~al.,}{{Serenelli} et~al.}{2021}]{Serenelli}
{Serenelli} A.,  et~al., 2021, \mn@doi [\aapr] {10.1007/s00159-021-00132-9}, \href {https://ui.adsabs.harvard.edu/abs/2021A&ARv..29....4S} {29, 4}

\bibitem[\protect\citeauthoryear{{Soydugan}, {Soydugan}  \& {Ali{\c{c}}avu{\c{s}}}}{{Soydugan} et~al.}{2020}]{Soydugan}
{Soydugan} F.,  {Soydugan} E.,   {Ali{\c{c}}avu{\c{s}}} F.,  2020, \mn@doi [Research in Astronomy and Astrophysics] {10.1088/1674-4527/20/4/52}, \href {https://ui.adsabs.harvard.edu/abs/2020RAA....20...52S} {20, 052}

\bibitem[\protect\citeauthoryear{{Tasdemir} \& {Yontan}}{{Tasdemir} \& {Yontan}}{2023}]{Tasdemir_2023}
{Tasdemir} S.,  {Yontan} T.,  2023, \mn@doi [Physics and Astronomy Reports] {10.26650/PAR.2023.00001}, \href {https://ui.adsabs.harvard.edu/abs/2023PARep...1....1T} {1, 1}

\bibitem[\protect\citeauthoryear{{Timmes} \& {Swesty}}{{Timmes} \& {Swesty}}{2000}]{Timmes2000}
{Timmes} F.~X.,  {Swesty} F.~D.,  2000, \mn@doi [\apjs] {10.1086/313304}, \href {https://ui.adsabs.harvard.edu/abs/2000ApJS..126..501T} {126, 501}

\bibitem[\protect\citeauthoryear{{Torres}, {Andersen}  \& {Gim{\'e}nez}}{{Torres} et~al.}{2010}]{Torres}
{Torres} G.,  {Andersen} J.,   {Gim{\'e}nez} A.,  2010, \mn@doi [\aapr] {10.1007/s00159-009-0025-1}, \href {https://ui.adsabs.harvard.edu/abs/2010A&ARv..18...67T} {18, 67}

\bibitem[\protect\citeauthoryear{Townsend}{Townsend}{2024}]{SDK}
Townsend R.,  2024, MESA SDK for Linux, \mn@doi{10.5281/zenodo.10624843}, \url {https://doi.org/10.5281/zenodo.10624843}

\bibitem[\protect\citeauthoryear{{Tun{\c{c}}el G{\"u}{\c{c}}tekin}, {Bilir}, {Karaali}, {Plevne}  \& {Ak}}{{Tun{\c{c}}el G{\"u}{\c{c}}tekin} et~al.}{2019}]{Guctekin_2019}
{Tun{\c{c}}el G{\"u}{\c{c}}tekin} S.,  {Bilir} S.,  {Karaali} S.,  {Plevne} O.,   {Ak} S.,  2019, \mn@doi [Advances in Space Research] {10.1016/j.asr.2018.10.041}, \href {https://ui.adsabs.harvard.edu/abs/2019AdSpR..63.1360T} {63, 1360}

\bibitem[\protect\citeauthoryear{Virtanen et~al.,}{Virtanen et~al.}{2020}]{scipy}
Virtanen P.,  et~al., 2020, \mn@doi [Nature Methods] {10.1038/s41592-019-0686-2}, \href {https://rdcu.be/b08Wh} {17, 261}

\bibitem[\protect\citeauthoryear{{Wilson}}{{Wilson}}{1979}]{wd2}
{Wilson} R.~E.,  1979, \mn@doi [\apj] {10.1086/157588}, \href {https://ui.adsabs.harvard.edu/abs/1979ApJ...234.1054W} {234, 1054}

\bibitem[\protect\citeauthoryear{{Wilson}}{{Wilson}}{1990}]{wd3}
{Wilson} R.~E.,  1990, \mn@doi [\apj] {10.1086/168867}, \href {https://ui.adsabs.harvard.edu/abs/1990ApJ...356..613W} {356, 613}

\bibitem[\protect\citeauthoryear{{Wilson} \& {Devinney}}{{Wilson} \& {Devinney}}{1971}]{wd1}
{Wilson} R.~E.,  {Devinney} E.~J.,  1971, \mn@doi [\apj] {10.1086/150986}, \href {https://ui.adsabs.harvard.edu/abs/1971ApJ...166..605W} {166, 605}

\bibitem[\protect\citeauthoryear{{Yontan} \& {Canbay}}{{Yontan} \& {Canbay}}{2023}]{Yontan-Canbay_2023}
{Yontan} T.,  {Canbay} R.,  2023, \mn@doi [Physics and Astronomy Reports] {10.48550/arXiv.2310.13582}, \href {https://ui.adsabs.harvard.edu/abs/2023PARep...1...65Y} {1, 65}

\bibitem[\protect\citeauthoryear{{Yontan} et~al.,}{{Yontan} et~al.}{2022}]{Yontan_2022}
{Yontan} T.,  et~al., 2022, \mn@doi [\rmxaa] {10.22201/ia.01851101p.2022.58.02.14}, \href {https://ui.adsabs.harvard.edu/abs/2022RMxAA..58..333Y} {58, 333}

\bibitem[\protect\citeauthoryear{{Y{\"u}cel} \& {Bak{\i}{\c{s}}}}{{Y{\"u}cel} \& {Bak{\i}{\c{s}}}}{2022}]{Gokhan}
{Y{\"u}cel} G.,  {Bak{\i}{\c{s}}} V.,  2022, \mn@doi [\mnras] {10.1093/mnras/stac2293}, \href {https://ui.adsabs.harvard.edu/abs/2022MNRAS.516.2486Y} {516, 2486}

\bibitem[\protect\citeauthoryear{{van Hamme}}{{van Hamme}}{1993}]{wd4}
{van Hamme} W.,  1993, \mn@doi [\aj] {10.1086/116788}, \href {https://ui.adsabs.harvard.edu/abs/1993AJ....106.2096V} {106, 2096}

\bibitem[\protect\citeauthoryear{{van Hamme} \& {Wilson}}{{van Hamme} \& {Wilson}}{2003}]{wd5}
{van Hamme} W.,  {Wilson} R.~E.,  2003, in {Munari} U.,  ed.,  Astronomical Society of the Pacific Conference Series Vol. 298, GAIA Spectroscopy: Science and Technology. p.~323

\makeatother
\end{thebibliography}

\bsp	
\label{lastpage}
\end{document}